\documentclass{PoS}
\PoS{PoS(HEP2005)410}
\usepackage{vmargin,graphicx,epsfig,subfigure}

\RequirePackage{xspace}

\def\ra                 {\ensuremath{\rightarrow}\xspace}

\def\Bbar  {\kern 0.18em\overline{\kern -0.18em B}{}\xspace}
\def\Bz    {\ensuremath{B^0}\xspace}
\def\Bzb   {\ensuremath{\Bbar^0}\xspace}
\def\Dbar  {\kern 0.18em\overline{\kern -0.18em D}{}\xspace}
\def\Dz    {\ensuremath{D^0}\xspace}
\def\Dzb   {\ensuremath{\Dbar^0}\xspace}
\def\BzBzb {\ensuremath{B^0 {\kern -0.16em \Bzb}}\xspace}
\def\piz   {\ensuremath{\pi^0}\xspace}
\def\Bd    {\ensuremath{B_d}\xspace}

\def\Bs    {\ensuremath{B_s}\xspace}
\def\Bsb   {\ensuremath{\Bbar_s}\xspace}
\def\BsBsb {\ensuremath{B_s {\kern -0.16em \Bsb}}\xspace}

\def\fiOne  {\ensuremath{\phi_1}\xspace}
\def\fiTwo  {\ensuremath{\phi_2}\xspace}
\def\fiThree  {\ensuremath{\phi_3}\xspace}
\def\rhobar {\ensuremath{\overline \rho}\xspace}
\def\etabar {\ensuremath{\overline \eta}\xspace}

\def\invfb   {\ensuremath{\mbox{\,fb}^{-1}}\xspace}
\def\invpb   {\ensuremath{\mbox{\,pb}^{-1}}\xspace}
\def\invps   {\ensuremath{\mbox{\,ps}^{-1}}\xspace}

\def\BB    {\ensuremath{B\Bbar}\xspace} 
\def\nub      {\ensuremath{{\overline{\nu}}}\xspace}
\def\nunub      {\ensuremath{\nu{\overline{\nu}}}\xspace}
\def\ellell     {\ensuremath{\ell^+ \ell^-}\xspace}
\def\KS    {\ensuremath{K^0_{\scriptscriptstyle S}}\xspace} 
\def\KL    {\ensuremath{K^0_{\scriptscriptstyle L}}\xspace} 
\def\Dpm    {\ensuremath{D^\pm}\xspace}
\def\Dz    {\ensuremath{D^0}\xspace}
\def\Kbar  {\kern 0.2em\overline{\kern -0.2em K}{}\xspace}
\def\Kzb    {\ensuremath{\Kbar^0}\xspace}
\def\Kstarzb  {\ensuremath{\Kbar^{*0}}\xspace}
\def\Kstarm   {\ensuremath{K^{*-}}\xspace}

\def\Dmd{\ensuremath{{\rm \Delta}m_d}\xspace}
\def\Dms{\ensuremath{{\rm \Delta}m_s}\xspace}
\def\Dmq{\ensuremath{{\rm \Delta}m_q}\xspace}

\def\Vcd  {\ensuremath{|V_{cd}|}\xspace}
\def\Vtd  {\ensuremath{|V_{td}|}\xspace}
\def\Vus  {\ensuremath{|V_{us}|}\xspace}

\def\Vts  {\ensuremath{|V_{ts}|}\xspace}
\def\Vub  {\ensuremath{|V_{ub}|}\xspace}
\def\Vcb  {\ensuremath{|V_{cb}|}\xspace}
\def\Vtb  {\ensuremath{|V_{tb}|}\xspace}

\def\Vtq  {\ensuremath{|V_{tq}|}\xspace}

\def\stb{\ensuremath{\sin\! 2 \beta   }\xspace}
\def\stbeff{\ensuremath{\sin\! 2 \beta_{\rm{eff}}   }\xspace}
\def\sta{\ensuremath{\sin\! 2 \alpha   }\xspace}

\def\BToDK  {\ensuremath{B^- \ra D^{(*)0}K^{(*)-}}\xspace}
\def\BToDKDbK  {\ensuremath{B^- \ra D^{(*)0}K^{(*)-},\overline D^{(*)0}K^{(*)-}}\xspace}
\newcommand{\jprl}       [1]  {{Phys.\ Rev.\ Lett.\ {\bf #1}}}
\newcommand{\plb}       [1]  {{Phys.\ Lett.\ B~{\bf #1}}}  
\newcommand{\jprd}       [1]  {{Phys.\ Rev.\ D~{\bf #1}}}
\newcommand{\epjc}      [1]  {{Eur.\ Phys.\ Jour.\ C~{\bf #1}}}

\title{Heavy Flavours and CP violation }
\ShortTitle{Heavy Flavours and CP violation }

\author{Marie-H\'el\`ene Schune \\
Laboratoire de l'Acc\'el\'erateur Lin\'eaire\\
B.P. 34\\
91898 Orsay Cedex, France \\
E-mail \email{schunem@lal.in2p3.fr}}

\abstract{Recent results on Heavy Flavours and CP violation are presented. After a short introduction a taste of K and D 
results is given. In a third part results on \Vtd and \Vts are summaryzed including \BB mixing results 
and \Bd radiative decays. A summary of \Vcb and \Vub measurements is presented in the fourth part. In the next section CP
violation measurements in the \Bd sector are shown. Finally in the last part the overall status of the determination 
of the CKM matrix is presented both in the context of the Standard Model and in the context of New Physics.}
\FullConference{International Europhysics Conference on High Energy Physics\\

		 July 21st - 27th 2005\\

		 Lisboa, Portugal}

\begin{document}

\section{Introduction}

Accurate studies of charm and beauty hadrons allow to test the Standard Model
in the fermionic sector in particular for tests of the CP violation mechanism in the B sector, 
provide information on non perturbative QCD parameters 
which can be compared with lattice QCD calculations and open a window for searching 
for New Physics through loop processes. \\

In the Standard Model, weak interactions among quarks are encoded
in a 3 $\times$ 3 unitary matrix: the CKM matrix. The existence of 
this matrix conveys the fact that quarks,
in weak interactions, act as linear combinations 
of mass eigenstates \cite{ref:cabi,ref:km}. The general form of the CKM matrix is : 
\begin{equation}
V =
\left ( \begin{array}{ccc} 
V_{ud} ~~ V_{us} ~~ V_{ub} \\
V_{cd} ~~ V_{cs} ~~ V_{cb} \\
V_{td} ~~ V_{ts} ~~ V_{tb}
\end{array} \right ).
\end{equation}

The CKM matrix can be parametrised in terms of four free 
parameters which are measured in several physics processes. In the  Wolfenstein 
approximation, these  parameters are named: $\lambda$, A, 
$\rho$ and $\eta$ and the CKM matrix can be parametrised as : 
\begin{equation}
\begin{array}{cccc}
V_{CKM} =
&
\left ( \begin{array}{ccc}
1 - \frac{\lambda^{2}}{2} &                \lambda  &                    A \lambda^{3} (\rho - i \eta) \\
   - \lambda              &           1 - \frac{\lambda^{2}}{2}  &          A \lambda^{2}             \\
A \lambda^{3} (1 - \overline{\rho} -i \overline{\eta})  &       -A \lambda^{2}  &                      1
\end{array} \right ) & + O(\lambda^{4}).
\end{array}
\label{eq:eqw}
\end{equation}
 with $ \overline{\rho} = \rho ( 1-\frac{\lambda^2}{2} )~~~;~~~ \overline{\eta} = 
\eta ( 1-\frac{\lambda^2}{2} )$\cite{ref:BLO}. It is worthwhile noting that the expressions 
for \Vus and \Vcb are valid up to order $\lambda^7$ and $\lambda^8$ respectively. 
CP violation is accommodated in the CKM
matrix with a single parameter and its existence is related to $\overline{\eta} \neq 0$.
\\
From the unitarity of the CKM matrix ($V V^{\dag} = V^{\dag} V = 1 $), non diagonal elements of the
matrix products, corresponding to six equations relating its
elements, can be written. In particular, in transitions involving $b$ quarks, the scalar 
product of the third column with the complex conjugate of the first row must vanish:
\begin{equation}
V_{ud}^{\ast} V_{ub}~+~ V_{cd}^{\ast} V_{cb}~+~ V_{td}^{\ast} V_{tb}~=~0
\label{eq:triangle}
\end{equation}

\begin{figure}[hbt!]
\begin{center}
\includegraphics[width=5cm]{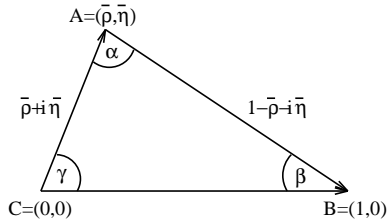}
\caption{The Unitarity Triangle.}
\label{fig:utria}
\end{center}
\end{figure}

This equation can be visualised as a triangle in the
($\overline{\rho},\overline{\eta}$) plane (Figure~\ref{fig:utria}).

The angles $\beta$ and $\gamma$ of the unitarity triangle are related
directly to the complex phases of the CKM-elements $V_{td}$ and
$V_{ub}$ respectively, through
\begin{equation}\label{e417}
V_{td}=|V_{td}|e^{-i\beta},\qquad V_{ub}=|V_{ub}|e^{-i\gamma}.
\end{equation}

Each of the angles is the relative phase of two adjacent sides :
\begin{eqnarray}
\beta  = & arg(\frac{V_{td}V_{tb}^*}{V_{cd}V_{cb}^*})\\
\gamma = & arg(\frac{V_{ud}V_{ub}^*}{V_{cd}V_{cb}^*})
\label{eq:utangle}
\end{eqnarray}

The angle $\alpha$ can be obtained through the relation $\alpha+\beta+\gamma=180^\circ$
expressing the unitarity of the CKM-matrix \footnote{There are two sets of notations for 
the angles of the Unitarity Triangle : 
$\alpha \equiv  $\fiTwo \ , $\beta \equiv$ \fiOne \ , $\gamma \equiv$ \fiThree .
Both will be used in the following.}.

The triangle shown in Figure \ref{fig:utria} 
-which depends on two parameters ($\overline{\rho},~\overline{\eta}$)-, 
plus $|V_{us}|$ and $|V_{cb}|$ give the full description of the CKM matrix. 

The Standard Model, with three families of quarks and leptons, predicts that all measurements 
have to be consistent with the point A($\overline{\rho},~\overline{\eta}$).
Extensions of the Standard Model can provide different predictions
for the position of the apex of the triangle, given by the 
$\overline{\rho}$ and $\overline{\eta}$ coordinates.\\
The most precise determination of these parameters is obtained using 
B decays, \Bz \-- \Bzb oscillations and CP asymmetry in the B and in the K sectors.
Many additional measurements of B meson properties (mass, branching fractions, lifetimes...) 
are necessary to constrain the Heavy Quark theories [Operator Product Expansion (OPE) /Heavy Quark 
Effective Theory (HQET) /Lattice QCD (LQCD)] to allow for precise extraction of the CKM parameters.  
\par
In principle {\it Heavy Flavours} deals with strange, charm and beauty hadrons. 
Due to lack of time, only a taste of K and D physics results will be given, emphasis will 
be put on B physics. Apologies to those whose work I did not have time to mention. 

\section{A taste of K and D results}
In this section emphasis will be given to new results related to CP violation. 
\subsection{Some K decays }
\begin{figure}[hbt!]
\begin{center}
{\includegraphics[height=6.0cm]{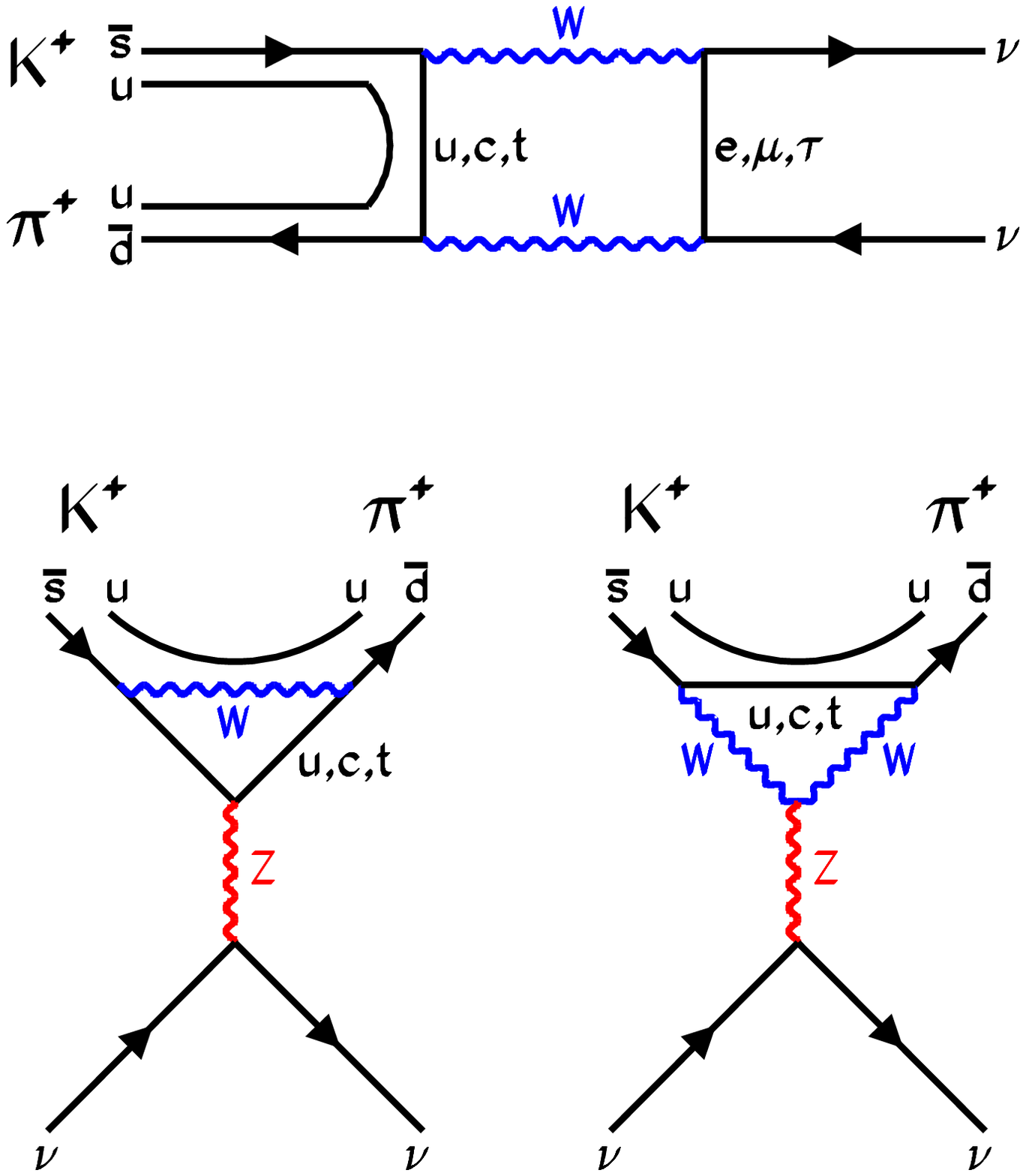}}
{\includegraphics[height=6.0cm]{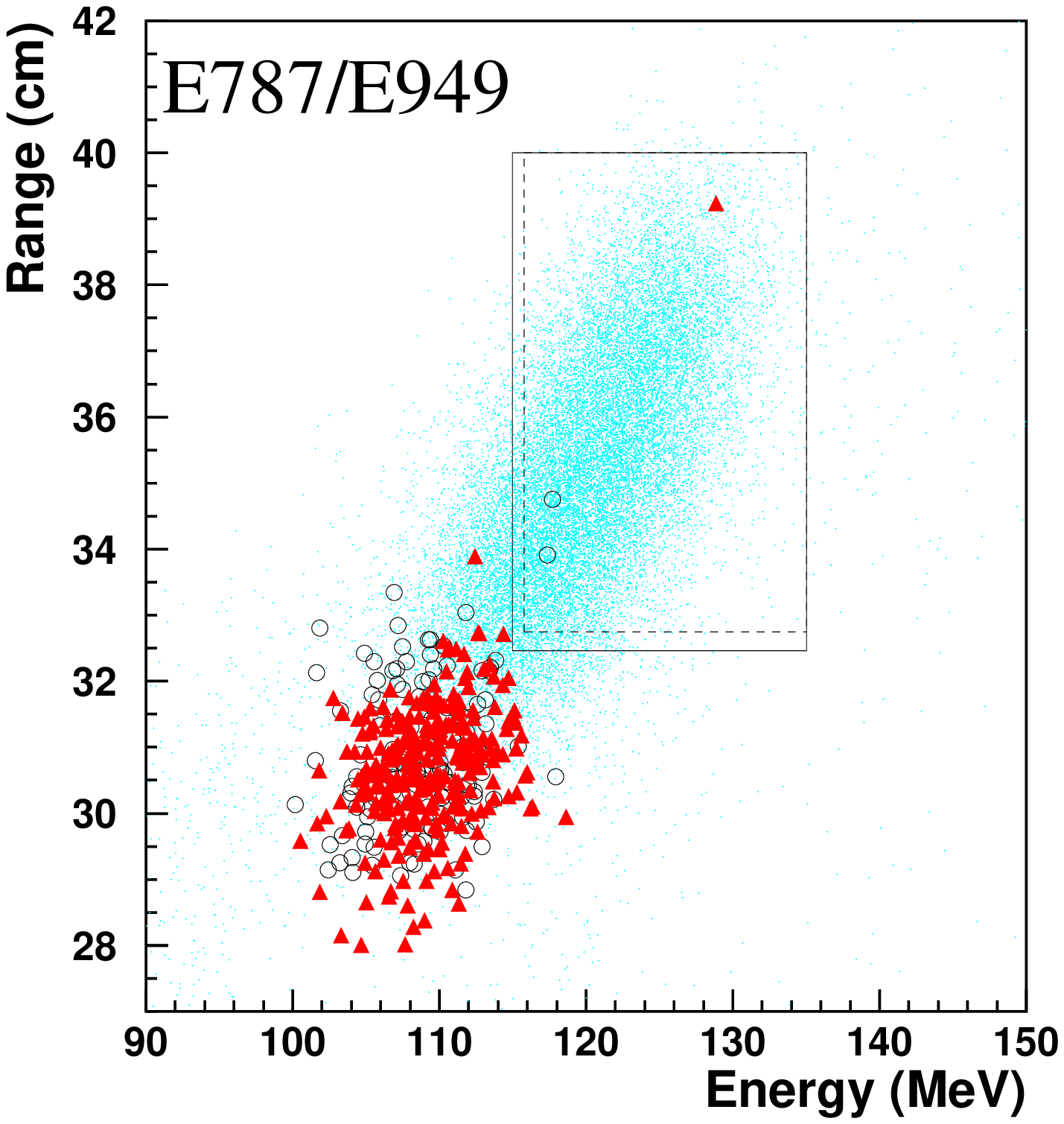}}
\caption{Left : the Feynman diagrams corresponding to the $K \ra \pi \nunub$ decay. Right : final plot of the 
E787/E949 experiments. The empty circles represent E787 data and the triangles E949 data.
The dots are signal Monte Carlo events. The solid (dashed) line box represents the signal region for E949 (E747). }
\label{fig:E949}
\end{center}
\end{figure}

The very rare decays $K \ra \pi \nunub$ (with branching fractions of the  order of $10^{-10}$ to $10^{-11}$) 
provide clean constraints on the CKM parameters but they 
are experimentally very challenging. Only, the charged decay  $K^{\pm} \ra \pi^{\pm} \nunub$ has been 
observed~\cite{bib:E787}, for the corresponding neutral one ($\KL \ra \piz \nunub$) only upper limits are available. 
The Feynmann diagram for the decay and the selection plot for $K^{\pm} \ra \pi^{\pm} \nunub$ are shown 
on Figure~\ref{fig:E949}. Other modes such as $K \ra \piz \ellell$ have been searched for, 
but only upper limits are available at present. The current status is summarised in Table~\ref{tab:rareKDecays}.

\begin{table}[htb]
\begin{center}
\begin{tabular}{|c|c|c|c|}
 \hline
Mode			      &  SM prediction		 & Exp. results	& CKM parameter \\ \hline
$K^{\pm} \ra \pi^{\pm} \nunub$&$(8.0 \pm 1.0) \ 10^{-11}$ & $1.47^{+1.30}_{-0.89} \ 10^{-10}$ \cite{bib:E787} E787/E949& $|\Vts^* \Vtd |$\\ 
\hline
$\KL \ra \piz \nunub$	      &$(3.0 \pm 0.6) \ 10^{-11}$ & $< 2.9 \ 10^{-7}$ \cite{bib:E391a} E391a& Im$(\Vts^* \Vtd)\sim \eta$\\
\hline
$\KL \ra \piz e^+ e^-$      &$(3.7 \pm 1.1) \ 10^{-11}$ & $< 2.8 \ 10^{-10}$ \cite{bib:KTeV-ee} KTeV& Im$(\Vts^* \Vtd)\sim \eta$\\
\hline
$\KL \ra \piz \mu^+ \mu^-$  &$(1.5 \pm 0.5) \ 10^{-11}$ & $< 3.8 \ 10^{-10}$ \cite{bib:KTeV-mumu} KTeV& Im$(\Vts^* \Vtd)\sim \eta$\\
\hline
 \end{tabular}
\caption{Summary of the current status for $K \ra \pi \nunub$ and $K \ra \piz \ellell$ decays. 
For the  $K \ra \piz \ellell$ decay modes, improvements on the theoretical uncertainty are expected. 
New Physics effects can be different for the electron and the muon channels. The upper limits are given at 90 \% CL.} 
\label{tab:rareKDecays}
\end{center}
\end{table}

\par
The branching fraction of the CP violating decay $\KS \ra \piz \piz \piz$ is expected to be of the order of $1.9 \ 10^{-9}$ 
in the Standard Model. The best limits obtained are summarised in Table~\ref{tab:K03pi0}. 
\begin{table}[htb]
\begin{center}
\begin{tabular}{|c|c|c|}
 \hline
Experiment		  &  Method					   & limit at 90 \% CL 	\\ 	\hline	 
NA48~\cite{bib:NA48-3pi0} &  $K^0$ beam : interference			   &   $< 7.4 \ 10^{-7}$  	\\ 	\hline	 
KLOE~\cite{bib:KLOE-3pi0} &  direct search (tagged \KS from $\phi$ decay)  &   $< 1.2 \ 10^{-7}$  \\ 	\hline	 
 \end{tabular}
\caption{Summary of the current status for the search of the CP violating decay $\KS \ra \piz \piz \piz$.} 
\label{tab:K03pi0}
\end{center}
\end{table}

\par 
Direct CP violation in the decay $K^{\pm} \ra \pi^{\pm} \pi^{\pm} \pi^{\mp}$ has been searched for using asymmetry 
in the comparison of the $K^+$ and the $K^-$ Dalitz plot. Standard Model expectations vary between $10^{-6}$ and
few $10^{-5}$, the NA48/2 collaboration has obtained a preliminary result consistent with no CP violation : 
$(0.5 \pm 3.8) \ 10^{-4}$~\cite{bib:NA48-DCPV}, improving by one order of magnitude previous results. 
\subsection{Leptonic and semi-leptonic charm decays}
\begin{figure}
\begin{center}
{\includegraphics[height=60mm]{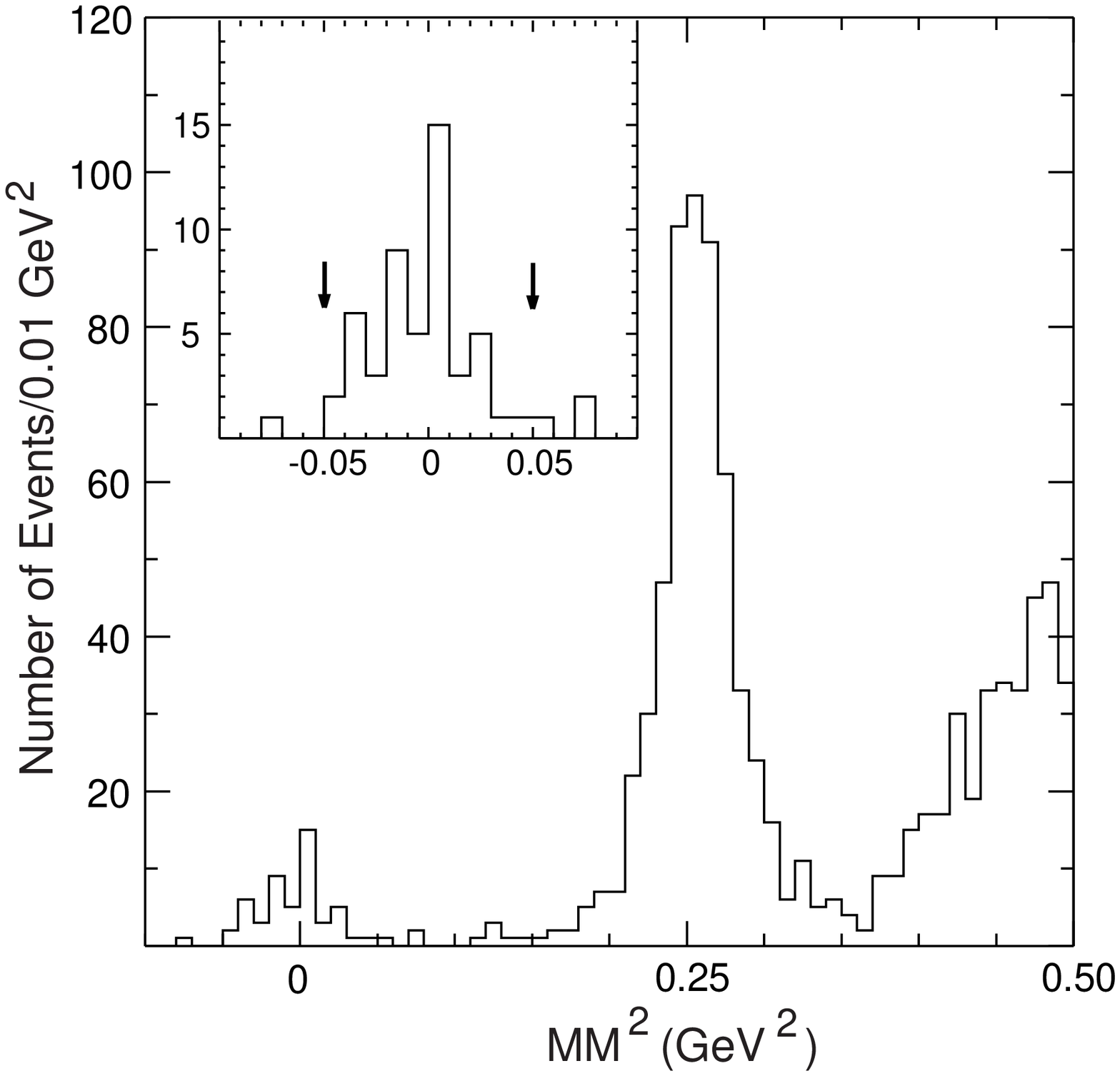}}
{\includegraphics[height=65mm]{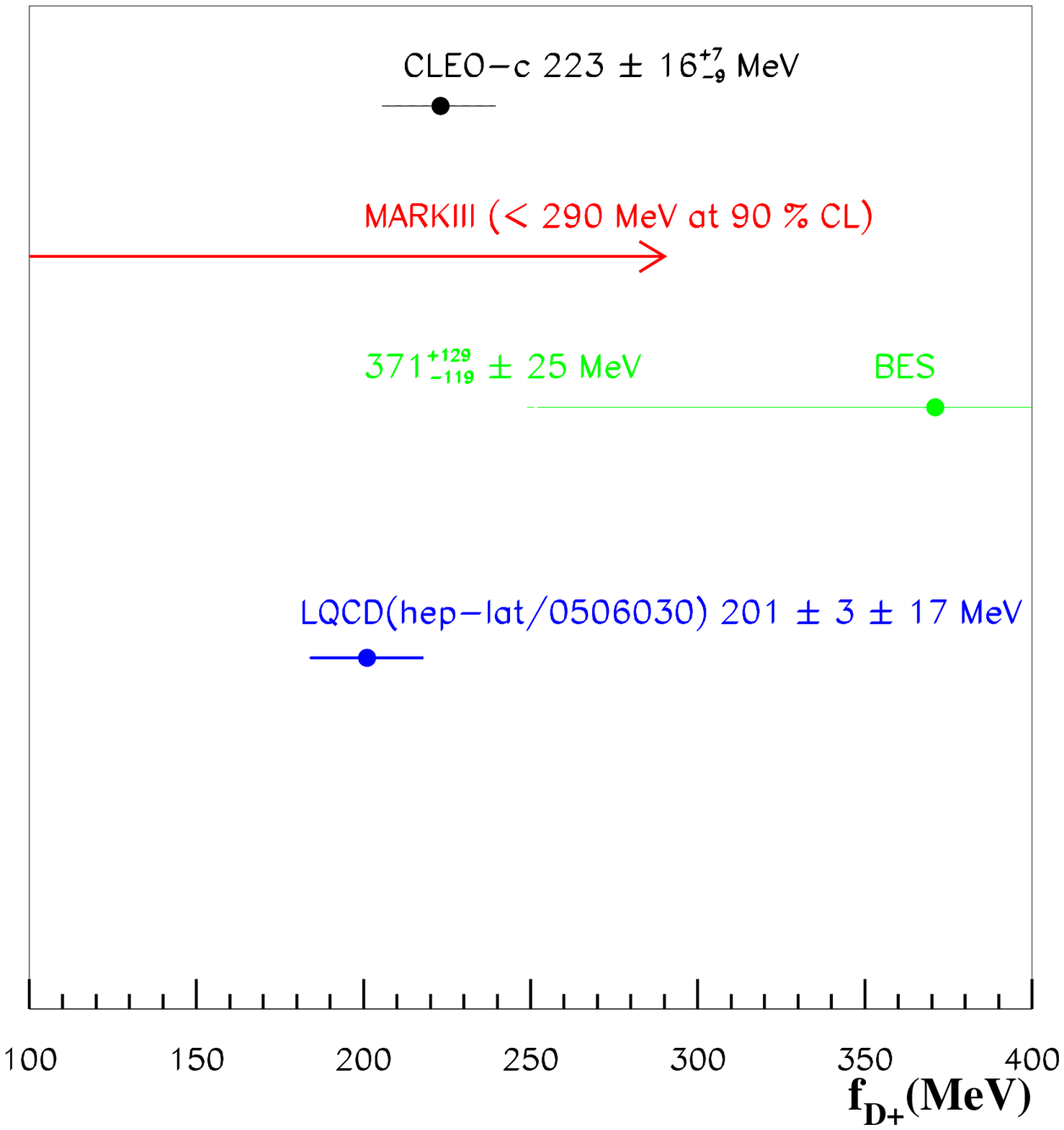}}
\caption{Left : the missing mass squared variable for the events passing all the selection cuts. 
The insert is a zoom on the signal region, the arrows indicate the cuts. There are 50 events in the signal region with a 
background of $2.93 \pm 0.50$ events. Right : comparison of the CLEO-c result for $f_D$ with previous experimental values 
and LQCD computation.}
\label{fig:cleoc-fD}
\end{center}
\end{figure}

The leptonic decay $D \ra \ell \nu$  width depends on few parameters : 
\begin{equation}
\Gamma (D\ra \ell \nu) = \frac{1}{8 \pi} G_F^2 f_D^2 m_{\ell}^2 M_D \left(1 -\frac{m_{\ell}^2}{M_D^2}\right)^2\Vcd^2
\end{equation}
Since the CKM matrix element \Vcd is precisely known the measurement of this partial width is equivalent to a measurement 
of $f_D$ the pseudo-scalar constant which translate the quarks wave functions overlap. It can be compared with theoretical 
predictions from non perturbative QCD calculations. The latest result has been obtained by 
the CLEO-c experiment which runs at the $\Psi^{''}(3770)$ resonance
decaying into a correlated $D \bar D$ pair. One charged $D$ is fully reconstructed (the tagging $D$), 
a muon  of charge opposite to the 
tagging D charge is searched for in the remaining tracks, requiring no additional activity in 
the calorimeters. 
The discriminating variable is the missing mass squared which should be compatible with
0 in case of signal (Figure~\ref{fig:cleoc-fD}). Using an integrated luminosity of 281 \invpb , a branching fraction of 
$(4.45 \pm 0.67 ^{+0.29}_{-0.36}) \ 10^{-4}$ is obtained~\cite{bib:cleoc-fD}. It can be translated into : 
$f_D = (223 \pm 16 ^{+7}_{-9})$ MeV, this result is compared with previous results and the latest LQCD computation 
(Figure~\ref{fig:cleoc-fD}). 

\par

Using the tagging D technique, various semi-leptonic decays of both \Dpm and \Dz have been reconstructed by CLEO-c.
They are compared with the present PDG values in Table~\ref{tab:cleoc-semilep}.   
\begin{table}[htb]
\begin{center}
\begin{tabular}{|c|c|c||c|c|c|}
 \hline
$D^+ \ra$ 				      &  CLEO-c (BF \%)			& PDG 	(BF\%)	&
$\Dz \ra$ 				      &  CLEO-c (BF \%)			& PDG 	(BF\%)		\\
\hline
$\Kzb e^+ \nu_e$	      & $8.71 \pm 0.38 \pm 0.37$	&$6.7 \pm 0.9$ & 
$ K^- e^+ \nu_e$	  	      & $3.44 \pm 0.10 \pm 0.10$	&$3.58 \pm 0.18$\\ \hline
$\piz e^+ \nu_e$	      & $0.44 \pm 0.06 \pm 0.03$	&$0.31 \pm 0.15$ & 
$ \pi^- e^+ \nu_e$	      & $0.262 \pm 0.025 \pm 0.008$	&$0.36 \pm 0.06$\\ \hline
$\Kstarzb e^+ \nu_e$	      & $5.56 \pm 0.27 \pm 0.23$	&$5.5 \pm 0.7$	& 
$ \Kstarm e^+ \nu_e$  	      & $2.16 \pm 0.15 \pm 0.08$	&$\--$\\ \hline
$\rho^0 e^+ \nu_e$	      & $0.21 \pm 0.04 \pm 0.01$	&$0.25 \pm 0.10$& 
$ \rho^- e^+ \nu_e$	      & $0.194 \pm 0.039 \pm 0.013$	&$\--$ \\ \hline
$\omega e^+ \nu_e$	      & $0.16 ^{+0.07}_{-0.06} \pm 0.01$	&$\--$ & & & \\ \hline
 \end{tabular}
\caption{Summary of the semileptonic decays modes branching fractions as measured by CLEO-c.} 
\label{tab:cleoc-semilep}
\end{center}
\end{table}
Since \Vcd is very well known, these measurements can be used to extract the D form factors, 
which can, in turn, be used in several ways as for example :  
\begin{itemize}
\item The form factor of the mode $D \ra K \ell \nu$ provide validation of LQCD computations. 
\item The form factors from $D \ra \rho / \pi  \ell \nu$ modes can be related to the B form factor for similar charmless 
decay modes, which helps reducing the theoretical uncertainty on the \Vub extraction.
\end{itemize}

\section{\Vtd and \Vts measurements}
The CKM matrix elements \Vtd and \Vts can be measured in the B physics sector 
through processes described by loop or box diagrams involving top quark contributions.
The presence of such diagrams allows also to search for New Physics since new particles 
may appear as well in the loops. 
\subsection{\BB mixing}
The probability that a meson \Bz produced at time $t=0$ transforms into a \Bzb (or stays as a \Bz) at time $t$ is given by :
\begin{eqnarray}
Prob(\Bz(t=0) \rightarrow \Bz(t) (\Bzb(t)))
        = \frac{1}{2} e^{- \Gamma t} ( 1 +(-) cos \Delta m t)
\label{eq:probMixing}
\end{eqnarray}
 In the Standard Model,  \BzBzb oscillations occur through a second-order
process -a box diagram- with a loop that contains W and up-type quarks. The box diagram with
the exchange of a $top$ quark gives the dominant contribution.
The oscillation probability is given in eq. (\ref{eq:probMixing}) and the time oscillation
frequency, which can be related to the mass difference between the light and heavy mass
eigenstates of the \Bd or \Bs meson system, is expressed in the SM, as : 
\begin{equation}
\Dmq\ = \frac{G_F^2}{6 \pi^2} \Vtb^2 \Vtq^2 M_W^2 M_{B_q} \eta_b S(x_t) f_{B_q}^2 \hat B_{B_q}
\label{eq:Dm}
\end{equation}
where $S(x_t)$ is the Inami-Lim function~\cite{bib:inami} and  $x_t=m_t^2/M^2_W$,
$m_t$ is the $\overline{MS}$ top quark mass, and
$\eta_b$ is the perturbative QCD short-distance NLO correction.
The remaining factor, $f_{B_q}^2 \hat B_{B_q}$, encodes the information
of non-perturbative QCD. Apart for the CKM matrix elements, the most uncertain parameter in this
expression is $f_{B_q} \sqrt{\hat B_{B_q}}$\footnote{The ratio 
$\xi=f_{B_s}\sqrt{\hat B_{B_s}}/f_{B_d}\sqrt{ \hat B_{B_d}}$ is expected to be better
determined from theory than the individual quantities entering into its expression.
The ratio $\Dmd/\Dms$ will thus be used to constrain the Unitarity Triangle.}.
\par
Experimentally the parameter \Dmd is now very precisely measured~\cite{bib:HFAG} : 
$\Dmd = 0.509 \pm 0.004 \invps$ ; the accuracy of the order
of 1 \%, is dominated by the B factories results.
\par
\begin{figure}[htpb]
\begin{center}
\includegraphics[width=90mm]{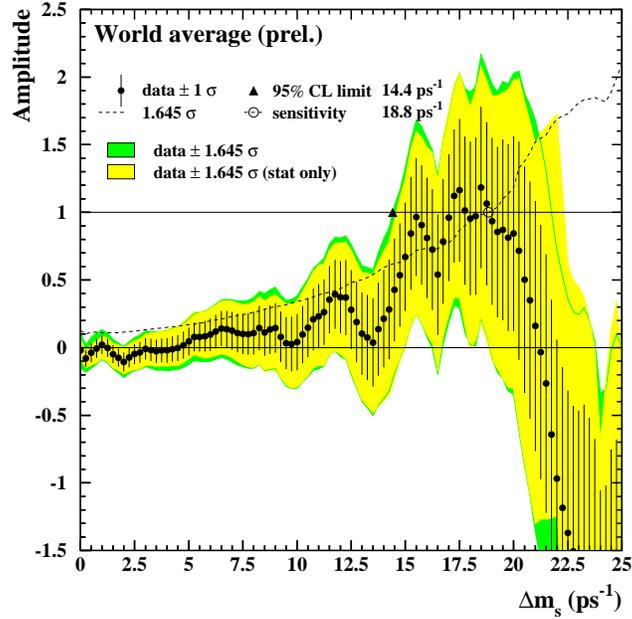}
\caption{The plot \cite{bib:HFAG} gives combined $\Delta {m}_s$ results from 
LEP/SLD/CDF analyses shown as an amplitude versus \Dms plot. 
The points with error bars are the data; the lines show the 95\% C.L. curves 
(darker regions correspond to the inclusion of systematics). 
The dotted curve corresponds to the sensitivity.}
\label{fig:dms_ampli}
\end{center}
\end{figure}

Due to the relative size of the CKM matrix elements the \Bs meson is expected to oscillate much faster 
than the \Bd meson and has not been measured yet.
The method used to set a limit on $\Delta {m}_s$ consists
in modifying Equation~\ref{eq:probMixing} in the following way \cite{bib:ampli}:
\begin{equation}
 1 \pm \cos{ (\Dms t)} \rightarrow 1 \pm {\cal A} \cos{( \Dms t)}. 
\label{eq:amplitude}
\end{equation}
${\cal A}$ and its error, $\sigma_{{\cal A}}$, are measured at fixed values  
of \Dms, instead of \Dms itself. 
In case of a clear oscillation signal, at a given frequency, the amplitude should be 
compatible with ${\cal A} = 1$ at this frequency.
With this method it is easy to set a limit. 
The values of \Dms excluded at 95\% C.L. are those satisfying the condition 
${\cal A}(\Dms) + 1.645  \sigma_{{\cal A}(\Dms)} < 1$. 
With this method, it is easy also to combine results from different experiments and to
treat systematic uncertainties in the usual way since, for each value of \Dms, 
a value for $\cal A$ with a Gaussian error $\sigma_{\cal A}$, is measured. 
Furthermore, the sensitivity of a given analysis can be defined as the value of
\Dms corresponding to 1.645 $\sigma_{\cal A} (\Dms) = 1$ (using 
${\cal A} (\Dms) = 0$), namely supposing that the ``true'' value of \Dms is well above the measurable value. 
Combining, with this amplitude method, the LEP and Tevatron 
results~\footnote{\Bs mesons are not produced at B factories.}\cite{bib:HFAG} one gets~:  
$\Dms  > 14.4 \ \invps$ at 95 \% C. L. with a sensitivity $\Dms = 18.8 ~\invps.$
The averaged amplitudes are shown in Figure~\ref{fig:dms_ampli}.   
\par
The only two running experiments which can study \BsBsb mixing today are D0 and CDF at Tevatron. 
Their current limits are summarised in Table~\ref{tab:Dms-Tevatron}. 
\begin{table}[htb]
\begin{center}
\begin{tabular}{|c|c|c|}
 \hline
Experiment							& Sensitivity		&    95 \% CL limit \\ \hline
CDF 355 \invpb ($D_s \ell \nu$ and $D_s \pi$) \cite{bib:Dms-CDF} 	& 8.4 \invps		& 	7.9 \invps  \\ \hline
D0 610  \invpb ($D_s \ell \nu$ ) 	 	 \cite{bib:Dms-D0}	& 9.5 \invps		& 	7.3 \invps  \\ \hline
 \end{tabular}
\caption{Summary of the Tevatron results on \Dms.}
\label{tab:Dms-Tevatron}
\end{center}
\end{table}
The experiments CDF and D0 have performed prospect studies, taking into account their current performances 
and foreseen experimental improvements~\cite{bib:Tevatron-Dms-Prospects}. Each experiment, 
with an integrated luminosity of about 4 \invfb 
(about 4 times the current one), should be able to push the \Dms limit up to about 20 \invps.   
\subsection{Radiative B decays}
\begin{figure}[htpb]
\begin{center}
\includegraphics[width=70mm]{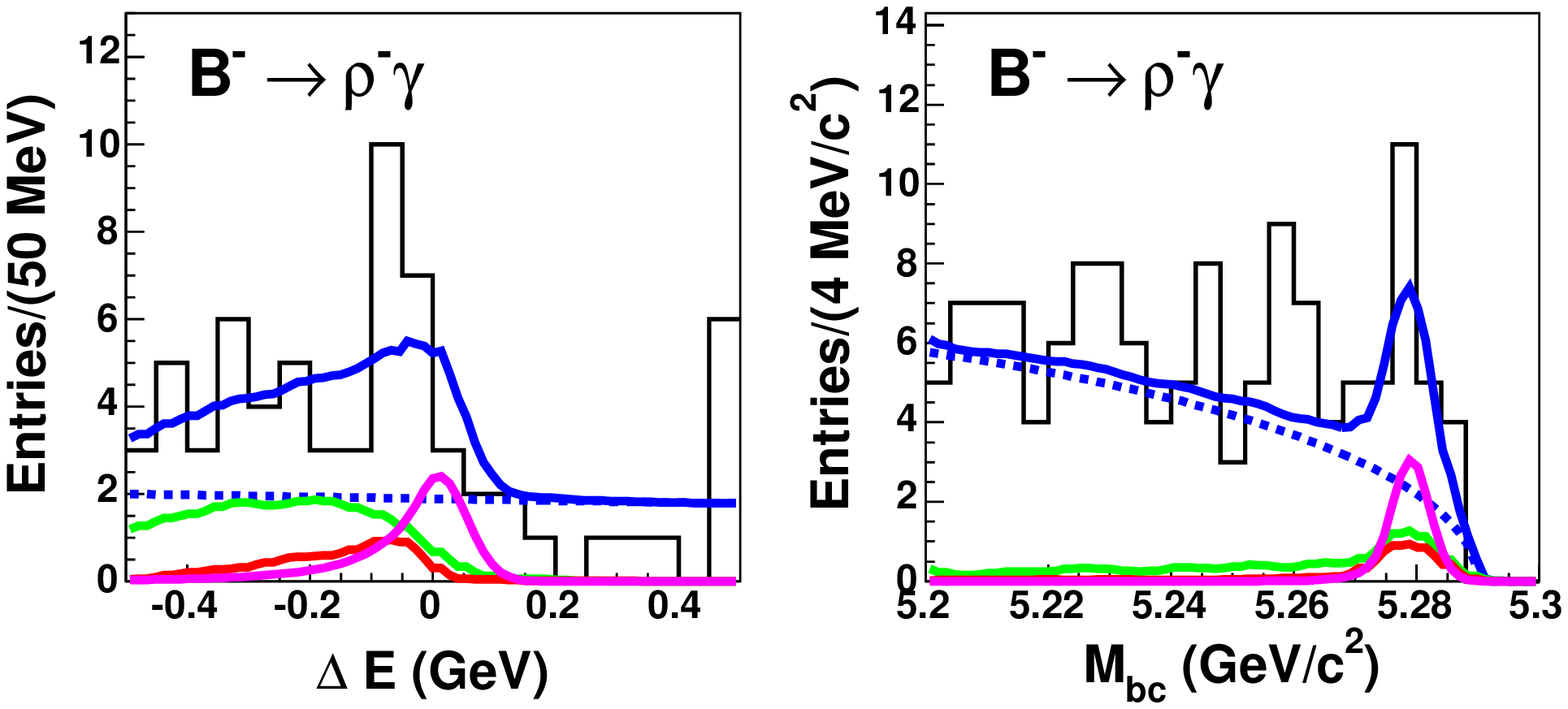}
\includegraphics[width=70mm]{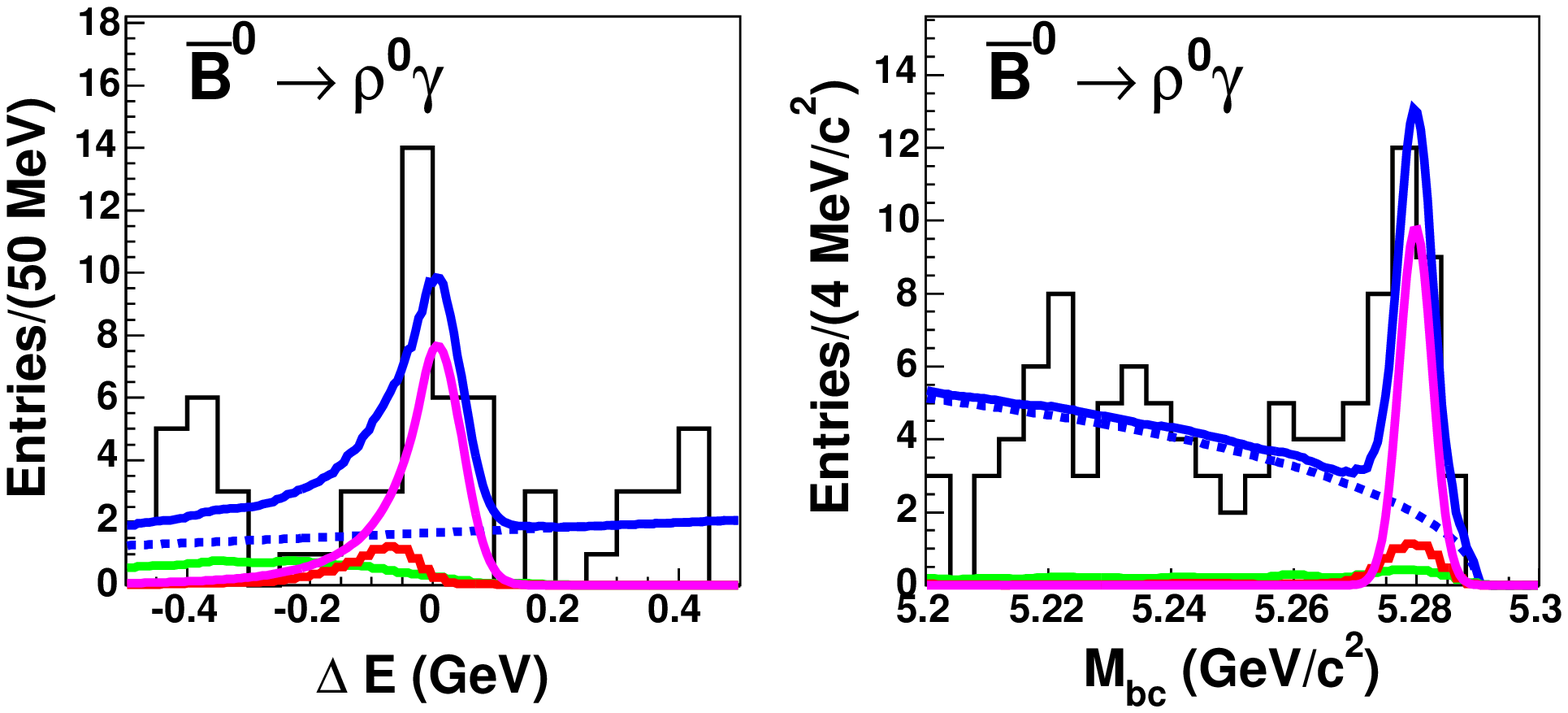}
\includegraphics[width=70mm]{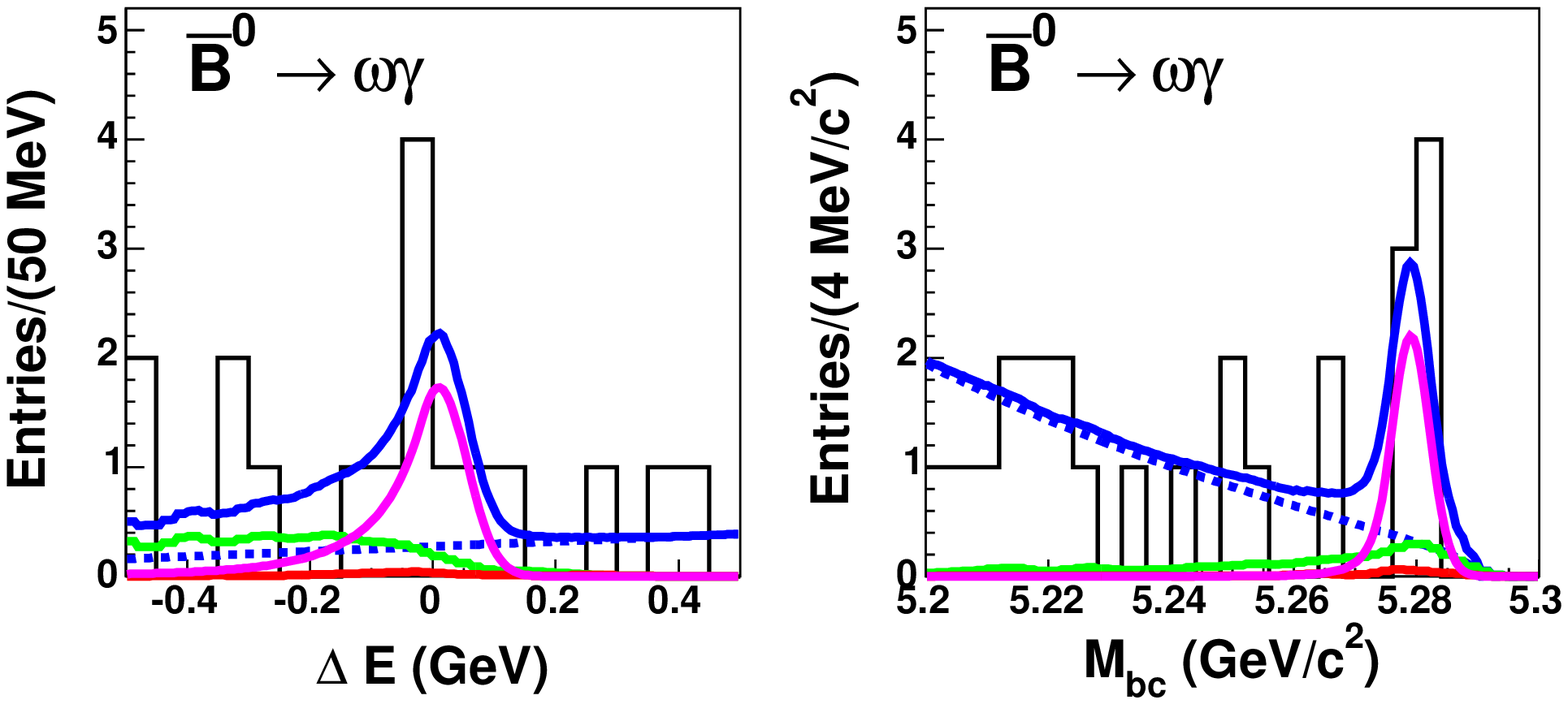}
\includegraphics[width=70mm]{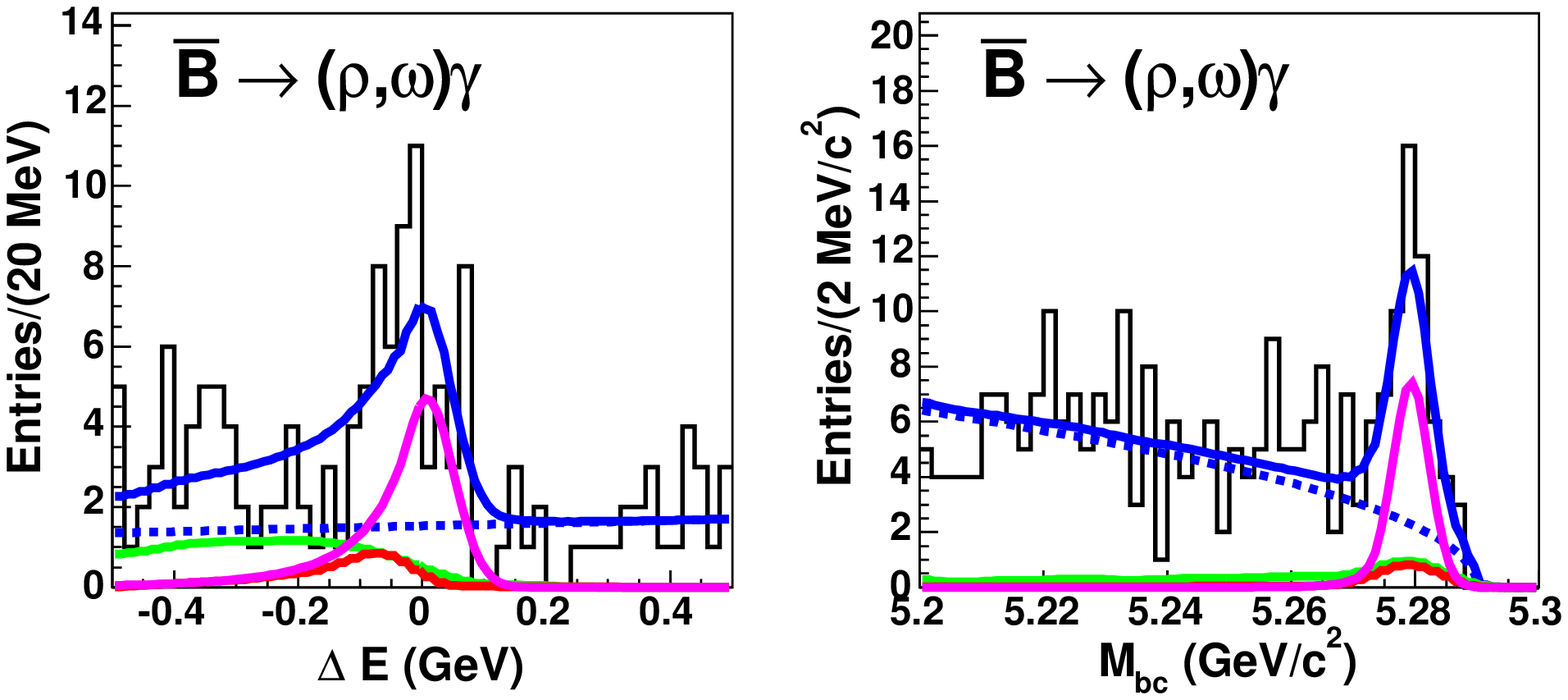}
\caption{Projections of the fit results~\cite{bib:BELLE-btodg} on the B mass and $\Delta E$ for the individual modes. 
Lines represent the signal (magenta), continuum (blue-dashed), $B \ra K^* \gamma$ (red), 
other $B$ decays background components (green) and the total fit result (blue-solid). The $\overline B$ symbol represent the sum 
for neutral and charged $B$ mesons decays.}
\label{fig:BELLE-btodg}
\end{center}
\end{figure}

Radiative B decays occur via penguins diagrams. Due to the difference in magnitude between the CKM matrix element 
\Vts and \Vtd, the $b \ra s \gamma$ type decays are much more copiously produced than the  $b \ra d \gamma$ type 
decays.
The study of the inclusive $\gamma$ energy spectrum in $b \ra s \gamma$ type decays 
gives information on the $b$ quark motion inside the $B$ meson and 
helps reducing the theoretical uncertainties in the \Vcb and \Vub extraction using semi-leptonic $B$ decays.
Two main types of analyses for the $b \ra s \gamma$ modes are performed~\cite{bib:btosg} :  fully inclusive 
ones where the photon energy spectrum 
is measured without reconstructing the $X_s$ system, and the backgrounds are suppressed using information 
from the rest of the event. The second method, the semi-inclusive one, uses a sum of exclusive final states 
in which possible $X_s$ system (about 55\% of the modes are reconstructed) are combined 
with the photon and kinematic constraints of the $\Upsilon(4S)$ 
reconstruction are used to suppress the background. 
The ratio of the $b \ra s \gamma$ and $b \ra d \gamma$ branching fractions 
could, in principle, provide information similar to the \BzBzb
mixing analysis. The theoretical uncertainties are smaller for the inclusive measurements, but this cannot be achieved 
for the $b \ra d \gamma$ decay due to the huge background. Only exclusive reconstruction can be performed for the time being. In that case the theoretical uncertainties are more difficult to control.
The first observation at $5.5 \sigma$ for these  $b \ra d \gamma$ type decays has been shown by BELLE. 
The results are summarised in Table~\ref{tab:RadPeng}. 
\begin{table}[htb]
\begin{center}
\begin{tabular}{|c|c|}
 \hline
Experiment	& BF($B \ra \rho / \omega \gamma$)		\\ \hline 
BABAR (211 $10^6$ \BB pairs) ~\cite{bib:BABAR-btodg} &  $< 1.2 \ 10^{-6}$ at 90 \% CL \\ \hline
BELLE (386  $10^6$ \BB pairs)~\cite{bib:BELLE-btodg} &  $(1.34^{+0.34 \ +0.14}_{-0.31 \ -0.10}) 10^{-6}$   \\ \hline
 \end{tabular}
\caption{Summary of the BABAR and BELLE results for the $b \to d \gamma$ analyses.}
\label{tab:RadPeng}
\end{center}
\end{table}
The selection plots for the BELLE analysis are shown in Figure~\ref{fig:BELLE-btodg}. 
\section{\Vcb and \Vub measurements}
\subsection{Semileptonic B decays}
The weak semi-leptonic decay of a free quark can be simply expressed : 
 \begin{equation}
	\Gamma_0 \equiv \Gamma ( b \ra q \ell \nub ) = \frac{G_F^2 \Vtq^2}{192 \pi^3} m_b^5 . 
 \end{equation}
However at the hadron level the expression gets more complicated due to the  hadronization 
process~\cite{bib:YellowBook}:
 \begin{equation}
	\frac{\partial^3 \Gamma}{\partial E_{\ell} \partial q^2 \partial m_X} = 
\Gamma_0 \times f(E_{\ell}, q^2, m_X) \times \left( 1+ \sum_n C_n \left( \frac{\Lambda_{QCD}}{m_b}\right)^n \right)
	\label{eq:semi-lept-q}
 \end{equation}
The term in parenthesis contains the perturbative and non-perturbative corrections to the semi-leptonic width. 
In exclusive decays it depends on QCD form factors which can be for example obtained from LQCD computations. 
In inclusive decays one uses Heavy Quarks symmetry and the OPE machinery. The OPE parameters can be obtained from data
using the spectra and moments from $b \ra s \gamma$ and $b \ra c \ell \nub$ distributions. 
In principle it works both for $b \ra c \ell \nub$ and $ b \ra u \ell \nub$ decays, however because of the very different 
values of the CKM matrix element one has to deal with $\Gamma( b \ra u \ell \nub)/\Gamma( b \ra c \ell \nub) \sim 50 $. 
Kinematical cuts have to be applied to reject this huge background, the measurements of the partial branching fractions
 will be performed in restricted phase space regions. As a consequence, the theoretical uncertainties will be more 
difficult to evaluate. 
\subsubsection{Extract of \Vub using inclusive semi-leptonic decays}
\begin{figure}[htpb]
\begin{center}
\includegraphics[width=5cm]{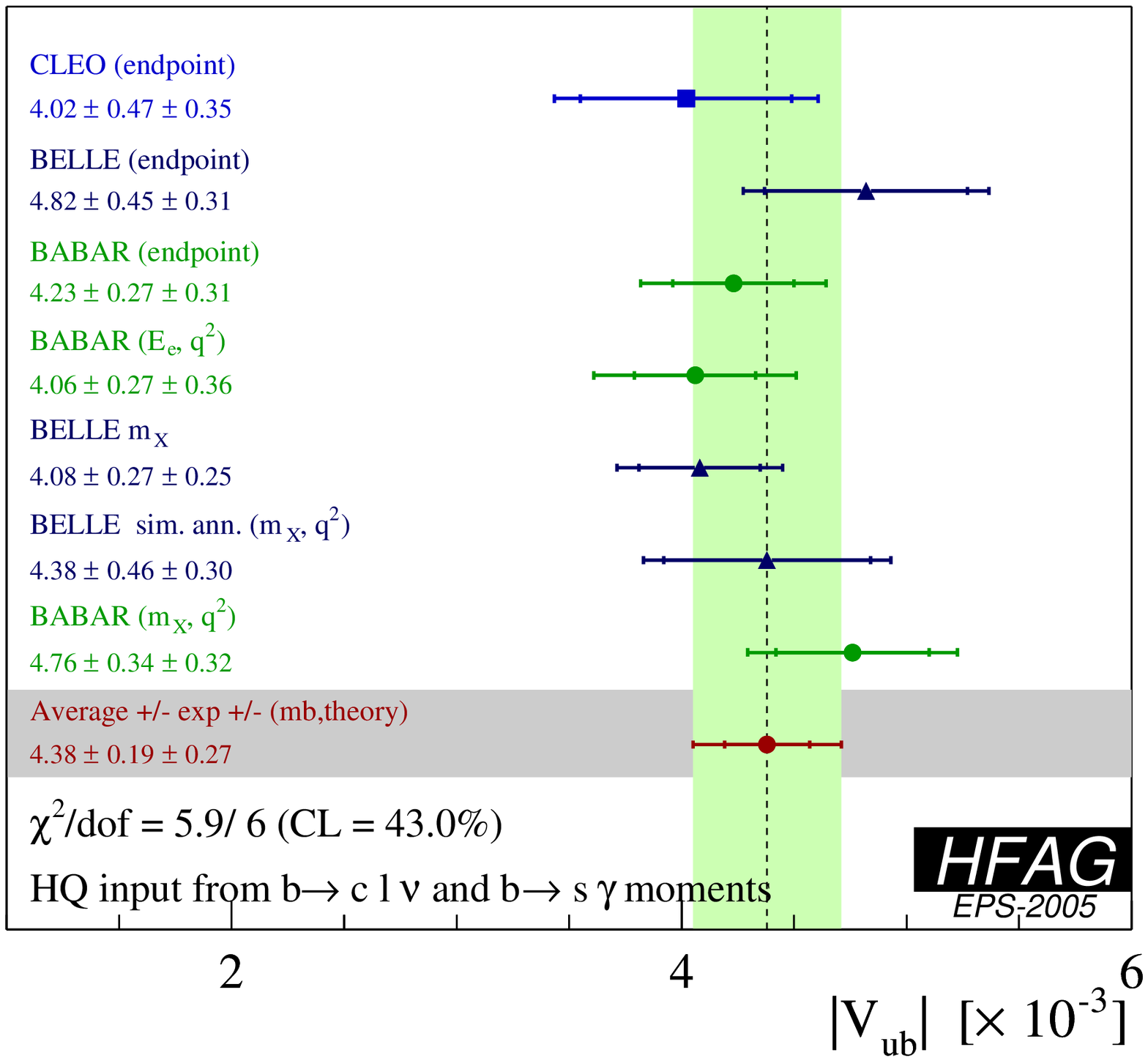}
\includegraphics[width=5cm]{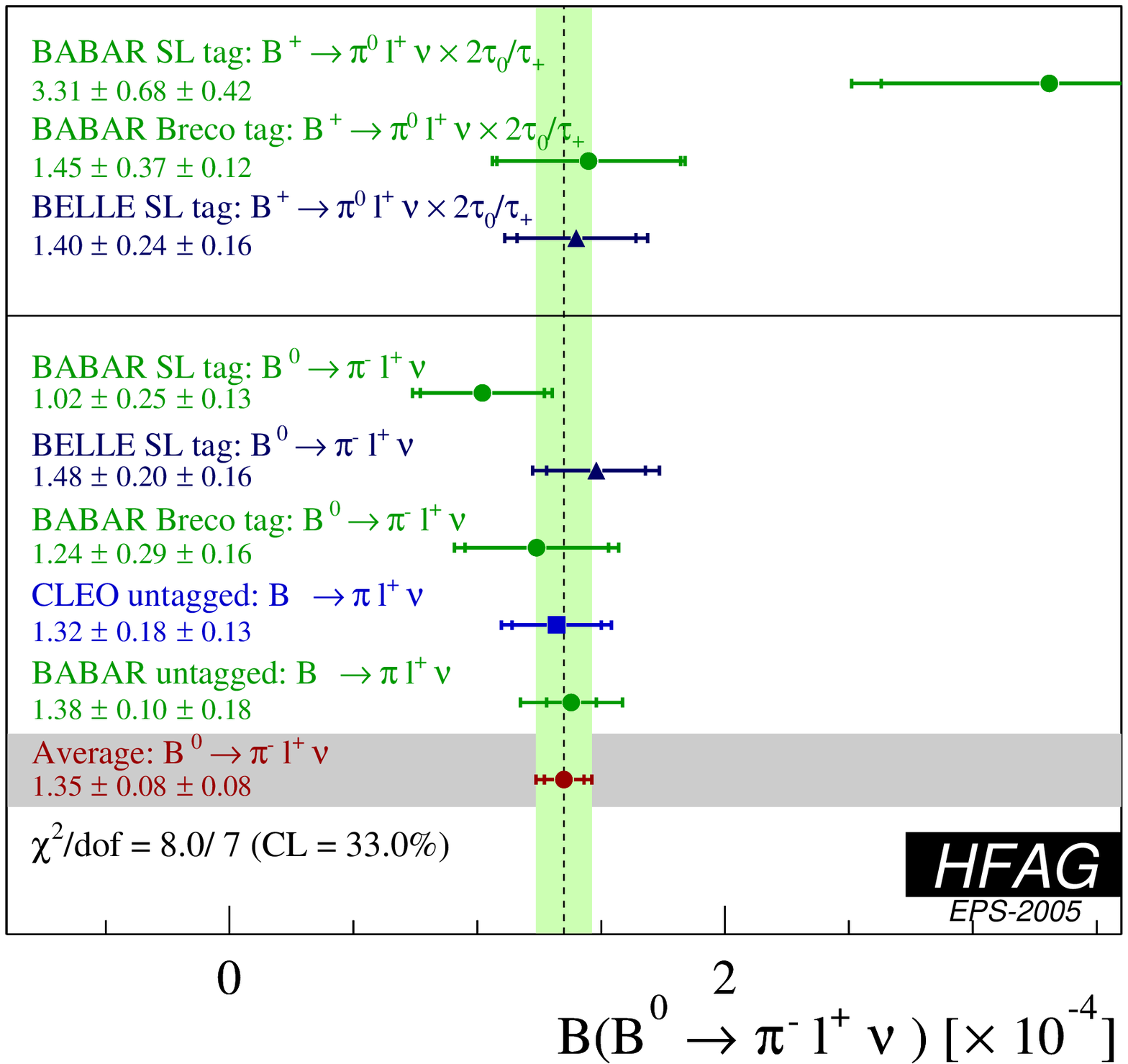}
\includegraphics[width=4cm]{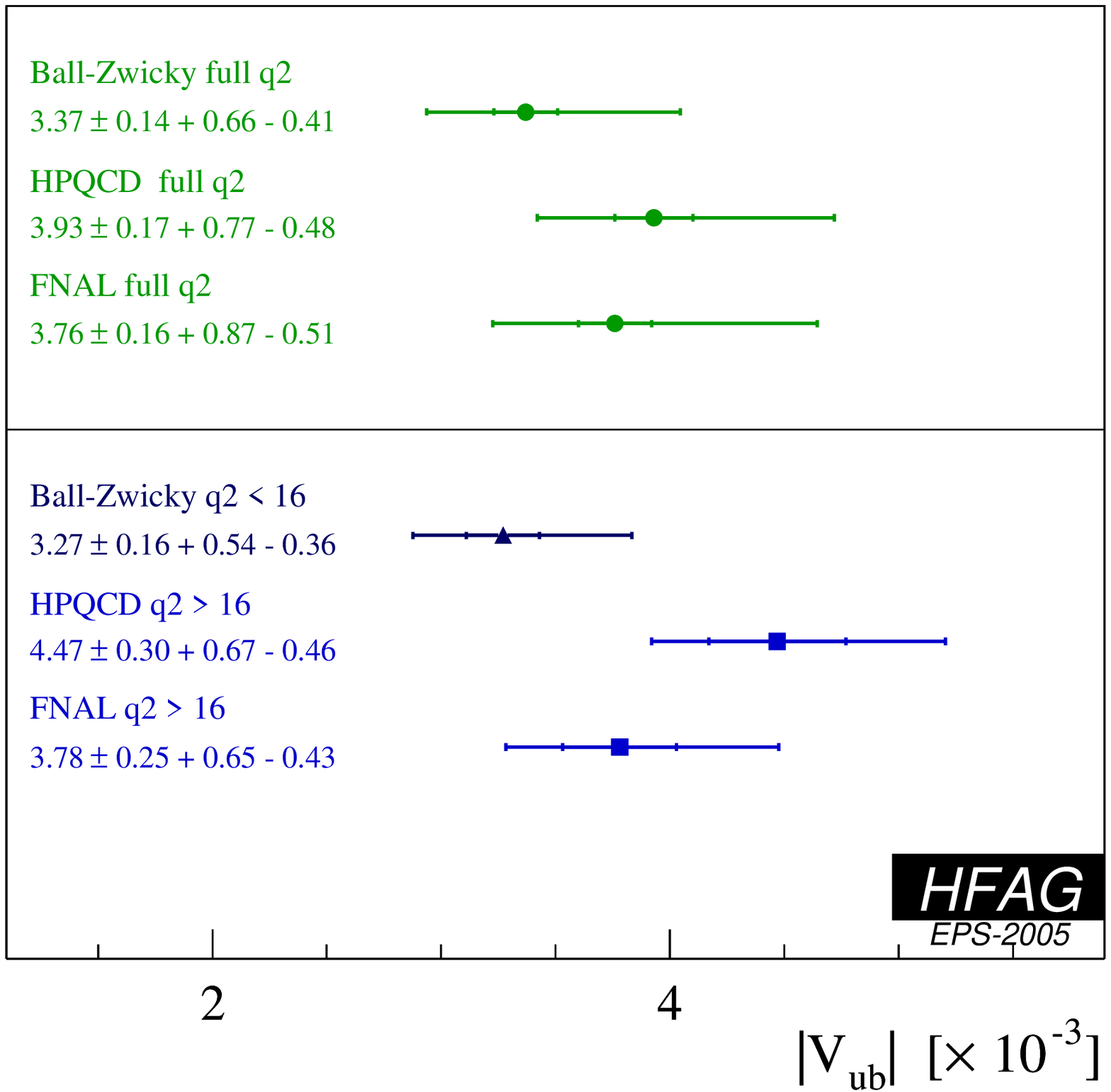}
\caption{Left plot : summary of  the \Vub values obtained from inclusive analyses. Middle plot : summary of the 
$B \ra \pi \ell \nub$ branching fraction measurements. Right plot : the \Vub values for different
FF computations.}
\label{fig:summaryVub}
\end{center}
\end{figure}

By using kinematical and topological variables, it is possible to 
select samples enriched in $b \ra u \ell^- \nub$ transitions.
There are, schematically, three main regions in the semi-leptonic 
decay phase space to be considered : the lepton energy end-point region where 
$E_{\ell}>\frac{M^2_B-M^2_D}{2M_B}$ (which was at the origin  for the first evidence of $b \rightarrow u$ transitions), 
the second region is the low hadronic mass region: $M_{X} < M_{D}$  
and the last one is the high $q^2$ region: $M^2_{\ell \nu}=q^2>(M_B-M_D)^2$
in which the background from $b \rightarrow c \ell^-  \overline{\nu_{\ell}}$ decays is small. In addition, in some cases one reconstructs (tags) the other B in order to improve the purity of the sample and to add additional kinematical constraints.
A summary of the various analyses~\cite{bib:vubinc} is given in Table~\ref{tab:Vub-inc} and the HFAG~\cite{bib:HFAG} average 
is shown on the right plot of Figure~\ref{fig:summaryVub}.

\begin{table}[htb]
\begin{center}
\begin{tabular}{|l|l|l|l|}
 \hline
Method				& Signal/Background	& Main points	  & \Vub $\times 10^{-3}$	\\ \hline
Untagged 			& 0.05 $\ra$ 0.2	& High statistics & $4.23 \pm 0.27_{\rm{exp}}\pm 0.31_{\rm{theo}}$\cite{bib:BABAR-El} \\ 
$\ell$ spectrum end point	&			& Delicate background &$4.82 \pm 0.45_{\rm{exp}}\pm 0.31_{\rm{theo}}$\cite{bib:BELLE-El}\\
$E_{\ell}>1.9 GeV$		&			& subtraction	  & 	\\ \hline
Untagged			& $\sim 0.5$		&High statistics	& $4.06 \pm 0.27_{\rm{exp}}\pm 0.36_{\rm{theo}}$\cite{bib:BABAR-l-all} \\
$\nu$ reconstruction		&			&Lower syst. on SF		& \\
Uses $M_X$			& 			&Delicate background 			& \\ \hline
Tagged 				&  $\sim 2 $		& Low background 			& $4.76 \pm 0.34_{\rm{exp}}\pm 0.32_{\rm{theo}}$\cite{bib:BABAR-tagged} \\
$M_X$ versus $q^2$ analyses	&            		& Small syst. on SF  		& $4.08 \pm 0.27_{\rm{exp}}\pm 0.25_{\rm{theo}}$\cite{bib:BELLE-tagged} \\
				& 			& Small statistics			& \\ \hline
 \end{tabular}
\caption{Summary of the inclusive analyses for the \Vub measurement.}
\label{tab:Vub-inc}
\end{center}
\end{table}

\subsubsection{Extract of \Vub using exclusive semi-leptonic decays}
\begin{figure}[htpb]
\begin{center}
\includegraphics[width=5cm]{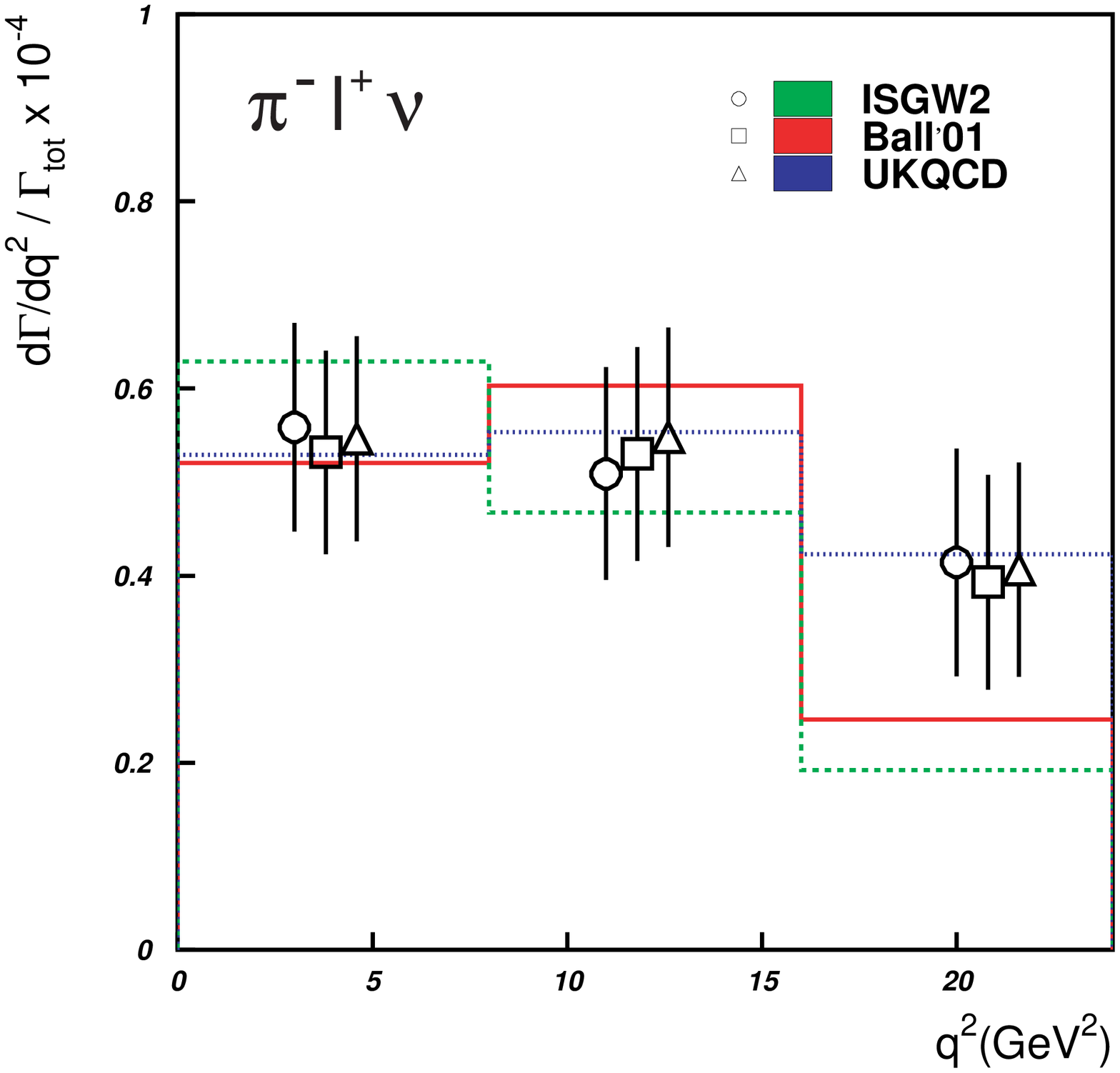}
\includegraphics[width=5cm]{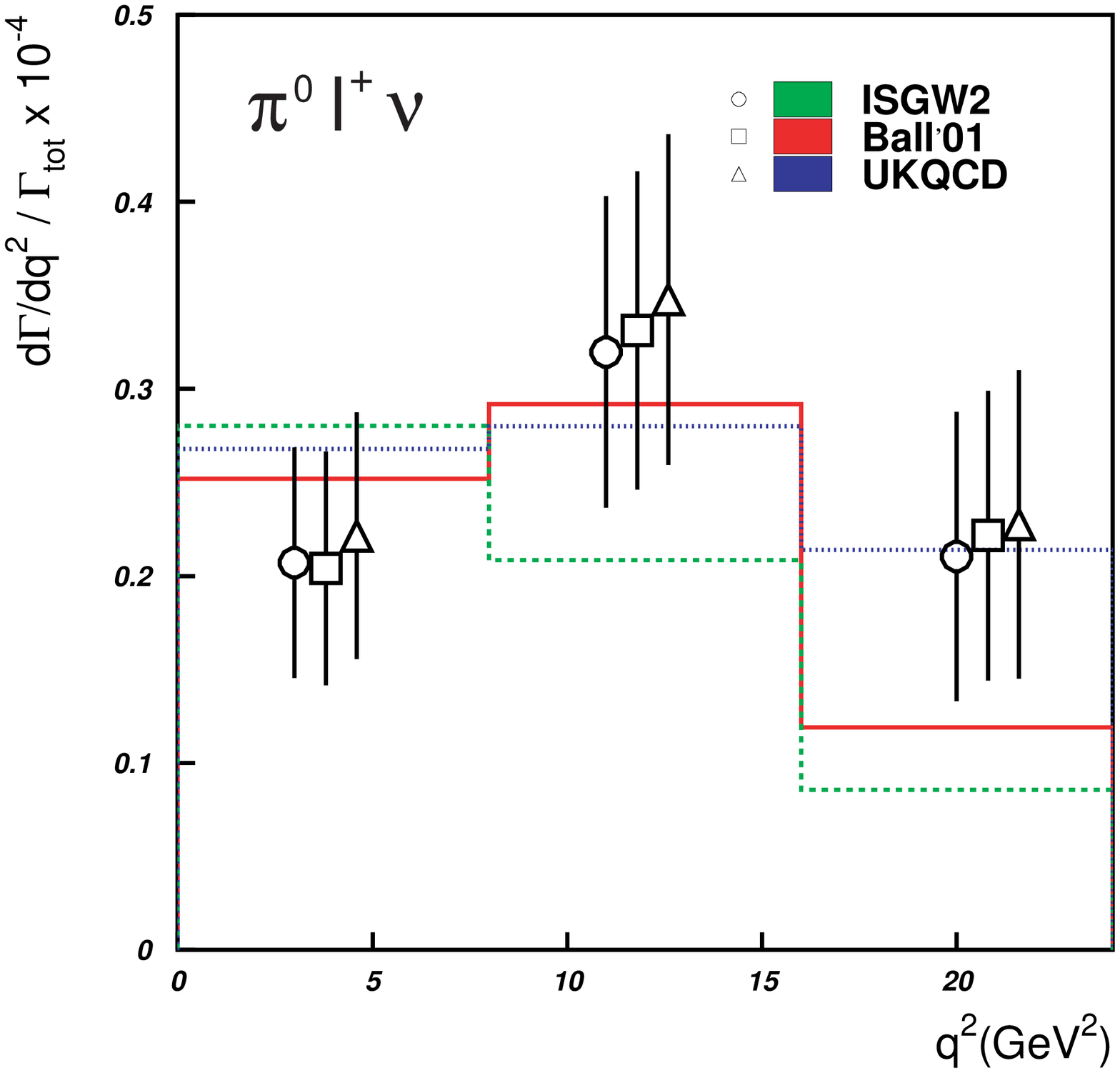}
\caption{Extracted $q^2$ distribution for the $B \ra \pi \ell \nub$ 
modes~\cite{bib:BELLE-Vub-q2}. Data points are 
shown for different Form Factor models used to estimate the detection efficiency. 
Lines are for the best fit of the Form factors shapes to the obtained $q^2$ distribution.}
\label{fig:BELLE-Vub-q2}
\end{center}
\end{figure}

The second method to determine \Vub consists in the reconstruction 
of charmless semileptonic B decays: $B \ra \pi (\rho) \ell \nub$.\\
The probability that the final state quarks  form a given meson is
described by form factors and, to extract \Vub from actual measurements, 
the main problem rests in the determination of these hadronic form factors.
Several theoretical approaches are used to compute these hadronic form factors. 
Experimentally one starts to be able to extract the signal 
rates in three independent regions of $q^2$. In this way it is possible 
to discriminate between models. An example is given  in Figure~\ref{fig:BELLE-Vub-q2} which
 shows that the ISGW~II model is only marginally compatible with the data.
This approach
could be used, in future, to reduce the importance of theoretical errors, 
considering that the ISGW ~II gave, at present, the further apart \Vub determination.
The summary of the  $B \ra \pi \ell \nub$ branching fraction measurements~\cite{bib:HFAG}
is given in the middle plot of Figure~\ref{fig:summaryVub} and can be translated into a \Vub measurement. 
There exist several theoretical computations of the Form Factors leading to different \Vub values, as can be seen 
from the right plot of  Figure~\ref{fig:summaryVub}.
Despite this precise measurement (of the order of 8 \%) the uncertainty on  \Vub is still of the order of 20\%, dominated
by theoretical uncertainty.  

The determination of \Vub from inclusive and exclusive semi-leptonic B decays are in agreement. The inclusive determination 
is the most precise one.
\subsubsection{\Vcb determination}
No new experimental results on \Vcb extraction were presented at this conference. The average for the inclusive determination 
is equal to $(41.58 \pm 0.45 \pm 0.58_{\Gamma_{SL}}) \ 10^{-3}$~\cite{bib:Vcb-inc} and the average for 
the exclusive method is : 
$\Vcb = (41.3 \pm 1.0 \pm 1.8) \ 10^{-3}$~\cite{bib:HFAG}. The two determinations are in good agreement. 
\subsection{$B \ra \tau \nu$}
The partial width of the $B \ra \tau \nub$ decay depends on few parameters : 
\begin{equation}
\Gamma (B\ra \tau \nu) = \frac{1}{8 \pi} G_F^2 f_B^2 m_{\tau}^2 M_B \left(1 -\frac{m_{\tau}^2}{M_D^2}\right)^2\Vub^2
\end{equation}
A measurement of this partial width is thus equivalent, in the Standard Model, to a measurement of $f_B \Vub$. Using
the value of \Vub from semi-leptonic decays and assuming the Standard Model, this could be translated 
into a $f_B$ measurement which could be compared 
with LQCD computations. In case of New Physics , there could be additional diagram with a  charged Higgs, such an 
analysis provides constraints on New Physics. Experimentally, in order to reduce the very 
large background, one B is fully reconstructed and the decay of interest 
is searched in the rest of the event. It is characterised by the presence of two neutrinos. 
The current results are summarised in Table~\ref{tab:BToTauNu}. The present limits are getting close to the Standard 
Model expected value ($8.2^{+1.7}_{-1.3}) \ 10^{-5}$ predicted by~\cite{bib:CKMFitter},\cite{bib:UTFit}. 
\begin{table}[htpb]
\begin{center}
\begin{tabular}{|l|l|}
 \hline
Experiment	& BF($B \ra \tau  \nu$) limit at 90 \% CL )		\\ \hline 
BABAR (232 $10^6$ \BB pairs))~\cite{bib:BABAR-BToTauNu}  &  $2.6 \ 10^{-4}$ \\ \hline
BELLE (275  $10^6$ \BB pairs)~\cite{bib:BELLE-BToTauNu} &  $1.8 \ 10^{-4}$   \\ \hline
 \end{tabular}
\caption{Summary of BF($B \ra \tau  \nu$) 90 \% CL limits.}
\label{tab:BToTauNu}
\end{center}
\end{table}
\section{CP violation in B decays}
Following~\cite{bib:Nir-PDG}, CP violation can be categorised into three types : 
\begin{description}
\item [CP violation in the decay] : it is the case where 
${\cal A}(B \ra f ) \neq  {\cal A}(\overline B \ra \overline f )$. There should exist
two amplitudes with different weak and strong phases to reach the final state $f$. 
This type of CP violation can be seen both in charged and neutral $B$ decays. 
\item [CP violation in the mixing] : it is the case where 
${\cal A}(\Bz \ra \Bzb ) \neq  {\cal A}(\Bzb \ra \Bz )$. This type of CP violation is 
due to the fact that the CP eigenstates are not the mass eigenstates.
\item [CP violation in the interference between mixing and decay] : it is due to 
the interference between a decay without mixing, $\Bz \ra f$ and a decay with mixing 
$\Bz \ra \Bzb \ra f$ (such an effect occurs only in decays where 
$f$ is common to \Bz and \Bzb). The most famous example 
is ${\cal A}(\Bz \ra J/\Psi \KS ) \neq  {\cal A}(\Bzb \ra J/\Psi \KS )$.
\end{description}
CP violation  has been first observed in the neutral Kaon system as the 
effect of CP violation in mixing. This type of CP violation is expected 
to be small ($10^{-3}$ to $10^{-4}$) in the \Bz meson
system. A large violation is possible in the Standard Model both as direct CP
violation and as time-dependent CP violation in the interference between mixing and
decay.
The time evolution of \Bz , taking into account CP violation can be written as :
\begin{equation} 
Prob(\Bz(t=0) \rightarrow \Bz(t) (\Bzb(t)))
        = \frac{1}{4 \tau} e^{- t/\tau} ( 1 +(-) C \cos (\Dmd t) -(+) S \sin (\Dmd t ) )
\label{eq:CP-Time}
\end{equation} 
The parameter $S$ is non-zero if there is mixing induced CP violation, while a non-zero 
value for $C$ would indicate direct CP violation. 

\subsection{$B \ra $ charmonium : $\beta$ or \fiOne}
For this type of decay the dominant penguin contribution has the same weak phase, so
no direct CP violation is expected to be seen. The only diagram with a different weak phase is
suppressed by a factor $\lambda^2$ and by OZI. For these $\Bz \ra (c \overline c) K^0$ decays
one should measure : $C=0$ and $S = \eta \stb$ with ($\eta = +1$ for \KS and $\eta = -1$ for \KL).      
The measurements of \stb ~\cite{bib:s2b} are summarised in Table~\ref{tab:s2b}. The overall average is 
$ \stb = 0.685 \pm 0.032$~\cite{bib:HFAG}, 
a 5 \% precision measurement. This precise value is in good agreement with the predicted one from fits using 
constraints only from sides of the  Unitarity Triangle~\cite{bib:CKMFitter},\cite{bib:UTFit}. 
This indicates a coherent description of CP violation within the Standard Model and that Standard Model is the dominant source of 
CP violation in the B meson sector.
\begin{table}[htb]
\begin{center}
\begin{tabular}{|l|l|l|}
 \hline
Experiment		  &   BABAR (227 $10^6$ \BB pairs)~\cite{bib:BABAR-s2b}	&  BELLE (386 $10^6$ \BB pairs)~\cite{bib:BELLE-s2b} \\ \hline
\stb from $c \overline c$ \KS & $0.75 \pm 0.04_{\rm{stat}}$			&  $0.668 \pm 0.047_{\rm{stat}}$ \\ \hline
\stb from $c \overline c$ \KL & $0.57 \pm 0.09_{\rm{stat}}$			&  $0.619 \pm 0.069_{\rm{stat}}$ \\ \hline
All charmonium 		      & $0.722 \pm 0.040_{\rm{stat}}\pm 0.023_{\rm{syst}}$&  $0.652 \pm 0.039_{\rm{stat}}\pm 0.020_{\rm{syst}}$ \\ \hline
 \end{tabular}
\caption{Summary of the \stb measurements.}
\label{tab:s2b}
\end{center}
\end{table}

\subsection{\BToDK : $\gamma$ or \fiThree \ , $\beta$ or \fiOne}
Different approaches have been used to measure the angle  $\gamma$ (or \fiThree )
of the Unitarity Triangle. They exploit the interference between $b \ra c$ and $b \ra u $ transitions.
Practically, this is achieved using decays of type \BToDKDbK\ with subsequent decays into final states 
accessible to both charmed meson and anti-meson. They are classified in three main types : 
\begin{description}
\item [The GLW method]~\cite{bib:GLW} : the \Dz meson decays into a CP final state
\item [The ADS method]~\cite{bib:ADS} : the \Dz meson is reconstructed into the $K \pi$ final state, 
for the $b \ra c$ ({\it resp.} $b \ra u$) transitions the \Dz decay mode will be the Cabbibo 
suppressed : $\Dz \ra K^+ \pi^-$ ({\it resp.} Cabbibo allowed : $\Dz \ra K^- \pi^+$) modes. 
In this way the magnitude of the two 
interfering amplitudes will not be too different.   
\item [The GGSZ method]~\cite{bib:GGSZ} : the \Dz final state is $\KS \pi \pi$ which is accessible to both 
\Dz and \Dzb. This requires analysis of the \Dz Dalitz plot, it can be seen as a mixture of the two previous
ones, depending on the position in the Dalitz plot. 
\end{description}
In all cases the involved diagrams are tree diagrams, so all methods should provide  
measurements of $\gamma$ independent of the possible existence of New Physics.
One of the main problem from the experimental point of view, is that the size of the CP
asymmetries involved depend on the ratio of the favoured and the \Vub and colour suppressed decays : 
$ r_B^{(*)} = |\frac{{\cal A}(B^- \ra \overline D^{(*)0} K^-}{\BToDK }|,$
which is, taking into account CKM matrix elements and colour suppression factors, expected to be of 
the order 0.1.  
An other experimental aspect is that the effective branching ratio is of the order of $10^{-7}$ in the case 
of the GLW and ADS methods. The situation is more favourable in the case of the GGSZ technique but this one is
complicated by the necessity to model the complex Dalitz plot of the $\Dz \ra \KS \pi \pi$ decay. 
Due to the very limited effective statistics and to the smallness of the 
$r_B$ parameter, the GLW and ADS methods are not yet able to measure $\gamma$~\cite{bib:Exp-GLW-ADS}.  
The results on $\gamma$ are summarised in Table~\ref{tab:gamma}. The large difference on the $\gamma$ statistical 
uncertainty between the BELLE and BABAR experiments cannot be attributed to the different sample sizes. It is 
rather due to different central values obtained for the various $r_B$ by the two experiments. 

\begin{table}[htb]
\begin{center}
\begin{tabular}{|l|l|l|l|}
 \hline
Exp				&	Mode	& $r_B$						    & $\gamma$	\\ \hline
BABAR\cite{bib:BABAR-gamma}	&$D K$		&$0.118 \pm 0.079 \pm 0.034 ^{+0.036}_{-0.034}$     & 	\\  
(227 $10^6$ \BB pairs) 		&$D^* K$	&$0.169 \pm 0.096 ^{+0.030 +0.029}_{-0.028 -0.026}$ & 	\\  
				&$DK^*$		&$0.05 \pm 0.11 \pm 0.05$			    &   \\
				& Combined	& 						    &$67 \pm 28 \pm 13 \pm 11^\circ$ \\ \hline
BELLE\cite{bib:BELLE-gamma}	&$D K$		&$0.21 \pm 0.08 \pm 0.03 \pm 0.04$ 		    & 	\\  
(275 $10^6$ \BB pairs)		&$D^* K$	&$0.12 \pm ^{+0.16}_{-0.11} \pm 0.02 \pm 0.04$	    & 	\\  
				& Combined	& 						    &$68 ^{+14}_{-15} \pm 13 \pm 11^\circ$ \\ 
				&$DK^*$		&$0.25 ^{+0.17}_{-0.18} \pm 0.09 \pm 0.09$	    &$112 \pm 35 \pm 9 \pm 11 \pm 8^\circ$ \\ \hline
 \end{tabular}
\caption{Summary of $r_B$ and $\gamma$ results. The last uncertainty is due to the \Dz Dalitz model parametrisation.}
\label{tab:gamma}
\end{center}
\end{table}

Despite the fact that the GLW and ADS analyses do not measure $\gamma$, they provide information on the $r_B$ parameters. 
All this information can be combined~\cite{bib:CKMFitter},~\cite{bib:UTFit}. The overall results from~\cite{bib:UTFit} are given in 
Table~\ref{tab:UTFit-rB-g}. From these numbers, it is clear that the $r_B$ parameters are rather on the low side so that the angle $\gamma$
will require large data samples to be measured. 

\begin{table}[htb]
\begin{center}
\begin{tabular}{|l|l|}
 \hline
$r_B(DK)   = 0.081 \pm 0.029$		& $r_B(DK) = 0.15 \pm 0.09$ \\ \hline
$r_B(D^*K) = 0.088 \pm 0.042$		& ${\bf \gamma = 66 \pm 17^\circ} $ \\ \hline 
 \end{tabular}
\caption{Summary of $r_B$ and $\gamma$ results given by~\cite{bib:UTFit} and taking into account BABAR and BELLE results from 
GLW, ADS and GGSZ methods.}
\label{tab:UTFit-rB-g}
\end{center}
\end{table}
\par
The analysis using the decay $\Bz \ra D^{(*)0} \pi^0 / \eta / \omega$ with $\Dz \ra \KS \pi \pi$ mode, 
which is similar to the GGSZ technique except that it requires a time dependent fit of the \Dz Dalitz plot density, provides
information on the angle $\beta$~\cite{bib:bondar}. This is important since with  the $\Bz \ra (c \overline c) K^0$ decays
one only measures \stb and an intrinsic ambiguity $2 \beta \leftrightarrow \pi - 2\beta$ remains. 
The BELLE collaboration has performed such an analysis and finds 
$\fiOne / \beta = (16 \pm 21 \pm 12)^\circ $~\cite{bib:BELLE-D0Pi0} 
which coincides with the Standard Model value of $\beta$ extracted from the \stb measurement. 
This is in agreement also with the result 
of~\cite{bib:cos2beta} which, using $\Bz \ra J/\Psi K^{*0}$ decays, shows that the  solution  $\cos (2 \beta) < 0$ 
is strongly disfavoured. 
\subsection{Charmless B decays : $\alpha$ or \fiTwo \ ,$\beta$ or \fiOne }
\subsubsection{The $ b \ra u \overline u d $ type transitions}
The decays of concern are $\Bz \ra \pi \pi$, $\rho \pi$ and $\rho \rho$
and follow the time dependence evolution of the \Bz meson of Equation~\ref{eq:CP-Time}.
Such decays suffer from the pollution of penguin contributions that, 
unlike the case of charmonium modes, do not have the same weak phase as the tree diagrams. 
If these penguins were negligible one would get $S = \sta$ and $C=0$. Since this is 
not the case one has $S=\sqrt{(1-C)}\sin 2 \alpha_{\rm{eff}}$ and the $C$ term is proportional to 
the relative penguin strong phase with respect to the tree amplitude. In order to extract 
$\alpha$ from $\alpha_{\rm{eff}}$ one will have to use theoretical arguments such as SU(2)-isospin. 
The $\Bz \ra \pi \pi$ results~\cite{bib:alpha-pipi} are summarised in Table~\ref{tab:alpha-pipi}, 
which shows that the discrepancy observed sometimes ago between BELLE and BABAR tends to disappear.
In order to extract $\alpha$ from these measurements one need the isospin related channel 
$B^{\pm} \ra \pi^{\pm} \pi^0$ and $\Bz \ra \pi^0 \pi^0$. Unfortunately, it turns out that the 
$\pi^0 \pi^0$ branching fraction is too small for a full isospin analysis but still visible, 
which is the sign that the penguin diagrams cannot safely be neglected. 
\begin{table}[htb]
\begin{center}
\begin{tabular}{|l|l|l|}
 \hline
		& $C_{\pi \pi}$			&      	$S_{\pi \pi}$  \\	\hline
BELLE		& $-0.56 \pm 0.12 \pm 0.06 $ 	&  $-0.67 \pm 0.16 \pm 0.06$   \\ \hline
BABAR		& $-0.09 \pm 0.15 \pm 0.04$	&  $-0.30 \pm 0.17 \pm 0.03$ \\ \hline
Average (HFAG)	& $-0.37 \pm 0.10$		& $-0.50 \pm 0.12$   \\ \hline
 \end{tabular}
\caption{Summary of the BELLE and BABAR results for $C_{\pi \pi}$ and $S_{\pi \pi}$ from ~\cite{bib:HFAG}.}
\label{tab:alpha-pipi}
\end{center}
\end{table}
A most favourable situation has been found with the mode  $\Bz \ra  \rho \rho$. In principle this channel 
requires an angular analysis of the final state, however it turns out that this decay is fully 
longitudinally polarised~\cite{bib:alpha-rhorho} and corresponds to a pure CP even state. 
The measured values for $C$ and $S$ are summarised in Table~\ref{tab:alpha-rhorho}. Contrary to the 
$ \pi \pi$ mode, the $\Bz \ra \rho^0 \rho^0$ decay has not been observed which indicates
a low penguin contamination. A useful bound on $|\alpha -\alpha_{\rm{eff}}|< 11 ^\circ$ can be 
obtained~\cite{bib:CKMFitter} which leads to : $\alpha = 96 \pm 13 ^\circ$. 

\begin{table}[htb]
\begin{center}
\begin{tabular}{|l|l|l|}
 \hline
		& $C_{\rho \rho}$			&      	$S_{\rho \rho}$  \\	\hline
BELLE		& $0.00 \pm 0.30 ^{+0.09}_{-0.10}$ 	&  $0.09 \pm 0.42 \pm 0.08$   \\ \hline
BABAR		& $-0.03 \pm 0.18 \pm 0.09$		&  $-0.33 \pm 0.24 ^{+0.08}_{-014}$ \\ \hline
Average (HFAG)	& $-0.030 \pm 0.17$			& $-0.21 \pm 0.22$   \\ \hline
 \end{tabular}
\caption{Summary of the BELLE and BABAR results for $C_{\rho \rho}$ and $S_{\rho \rho}$ from ~\cite{bib:HFAG}.}
\label{tab:alpha-rhorho}
\end{center}
\end{table}

Adding the additional constraints from the time dependent CP analysis of the $\Bz \ra \rho \pi$ decay 
mode (which helps in 
disfavouring the mirror solution at $\alpha +\pi/2$), one gets
a rather precise measurement : $\alpha = 99^{+12\ \circ}_{-9}$~\cite{bib:CKMFitter},\cite{bib:UTFit}. 
\par
Charmless B decays is also an active field of search for direct CP violation signals~\cite{bib:DCPV}. 
Despite important number of channels analysed, it is only seen with a significance greater than 
$4 \sigma$ in the $K \pi$ channel for two-body modes. For the three-body modes it has only been seen
at $3.9 \sigma$ in the $K^{\pm} \pi^{\pm} \pi^{\mp}$ channel by the BELLE collaboration~\cite{bib:BELLE-KPiPi}. 
In this last case a full Dalitz analysis is performed and direct CP violation is seen in the $K \rho^0$ channel. 
The results are summarised in Table~\ref{tab:DCPV}. 

\begin{table}[htb]
\begin{center}
\begin{tabular}{|l|l|l|}
 \hline
Experiment			& A($K^{\pm} \pi^{\mp}$)			& A($K^{\pm} \rho^0$)		\\ \hline
BABAR 	& $-0.133 \pm 0.030 \pm 0.009$ ~\cite{bib:BABAR-KPi}	& $0.34 \pm 0.13_{\rm{stat}} \pm 0.06_{\rm{syst}} \ ^{+0.15}_{-0.20 \rm{model}}$~\cite{bib:BABAR-KPiPi} \\
(227 $10^6$ \BB pairs)		&  				& 			\\ \hline
BELLE 	& $-0.113 \pm 0.022 \pm 0.008$~\cite{bib:BELLE-KPi} & $0.30 \pm 0.11_{\rm{stat}} \ ^{+0.11}_{-0.04 \rm{syst+model}}$~\cite{bib:BELLE-KPiPi} \\ 
 (386 $10^6$ \BB pairs)		& 				& 			\\ \hline
 \end{tabular}
\caption{Summary of the direct CP asymmetries observed in the $K^{\pm} \pi^{\mp}$ and $K^{\pm} \pi^{\pm} \pi^{\mp}$ modes.}
\label{tab:DCPV}
\end{center}
\end{table}
\subsubsection{The $ b \ra s \overline s s $-type transitions}
\begin{figure}[htpb]
\begin{center}
\includegraphics[width=8cm]{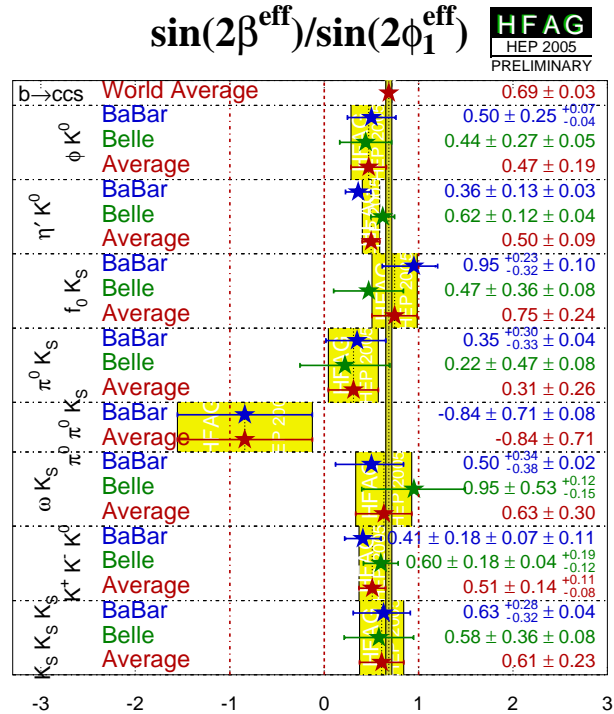}
\caption{Summary~\cite{bib:HFAG} of the BELLE and BABAR results for \stbeff obtained from  $ b \ra s \overline s s $-type transitions. 
The average value obtained for \stb from charmonium modes is also indicated.} 
\label{fig:s2beff}
\end{center}
\end{figure}

Example of such decays are $\Bz \ra \Phi \KS , \eta ' \KS , K^+ K^- \KS$ \dots 
These decays are due to loop diagrams and as such are sensitive to New Physics : additional 
diagrams with heavy particles in the loop and new CP violating phases may contribute to the decay amplitudes. 
The measurement of CP violation in these channels and the comparison with the reference value from charmonium 
modes is thus a sensitive probe of New Physics. Indeed, if no New Physics diagrams are present
the $S$ coefficient  in  $ b \ra s \overline s s $-type transitions should be close to \stb obtained from charmonium channels. 
Unfortunately, depending on the modes it is not exactly equal to \stb and the corrections are often difficult to compute. 
The cleanest (theoretically) mode $\phi \KS$  should lead to a $S$ parameter equal to \stb at the 5 \% level. 
Experimentally these modes are  more challenging than the charmonium ones due to smaller branching 
fractions and higher backgrounds. 
Many modes have been studied, BELLE and BABAR results are getting more accurate and in better agreement~\cite{bib:s2beff}. 
The results are summarised in Figure~\ref{fig:s2beff}. Several points are worthwhile to emphasise : 
\begin{itemize}
\item All modes (except $\eta ' \KS$ and $\pi^0 \pi^0 \KS$) are less than 
$1.5 \sigma$ away from the charmonium value.  
\item All the values for \stbeff modes are systematically below the \stb value from the charmonium modes
\item Recent QCD factorisations estimates~\cite{bib:s2beff-QCDF} point to $\stbeff > \stb$ and thus cannot explain the previous point.     
\item More statistics is needed in order to be able to conclude on this subject.
\end{itemize}
\section{Overall status}
\begin{figure}[htpb]
\begin{center}
\vskip 5mm
\includegraphics[width=10cm]{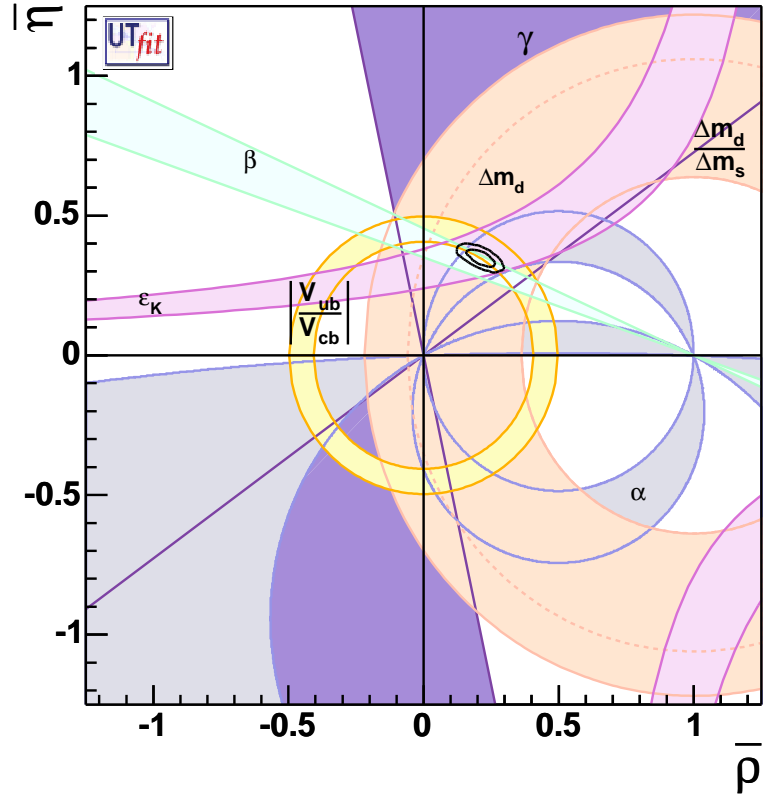}
\caption{Allowed regions for \rhobar and \etabar obtained by~\cite{bib:UTFit}. The closed contours at 68 \% and 95 \% 
probability are shown. The full lines correspond to 95 \% probability regions for the individual constraints, given 
by the measurements of \Vub/\Vcb , $\varepsilon_K$, \Dmd, \Dms, $\alpha, \beta, \gamma$. The dotted curve corresponds to the 95 \% upper limit 
obtained from the experimental study of \Dms.} 
\label{fig:UTFit-rhoeta}
\end{center}
\end{figure}

Global fits of the four CKM parameters taking into account 
the measurements of the three angles $\alpha, \beta$ and $\gamma$, \Vub 
and \Vcb CKM matrix elements, \Dmd and \Dms mixing frequencies and the direct CP violation parameter in the Kaon sector 
$\epsilon_K$) are performed~\cite{bib:CKMFitter}, 
\cite{bib:UTFit}. Besides slightly different theoretical inputs and different statistical treatments
both fitters agrees on the \rhobar and \etabar  values : 
\begin{eqnarray*}
		& \rhobar			& \etabar		\\
\rm {CKMFitter}	& 0.208^{+0.038}_{-0.043}	& 0.337^{+0.024}_{-0.022}	\\
\rm {UTfit}	& 0.216\pm 0.036		& 0.342 \pm 0.022 \\
\end{eqnarray*}
An example of a fit output is shown in Figure~\ref{fig:UTFit-rhoeta}. 
The fact that all measurements are compatible indicates that the Standard Model
provides a coherent picture of the CP violation mechanism and that New Physics should
appear as a correction to this framework.

\begin{figure}[htpb]
\begin{center}
\includegraphics[width=8cm]{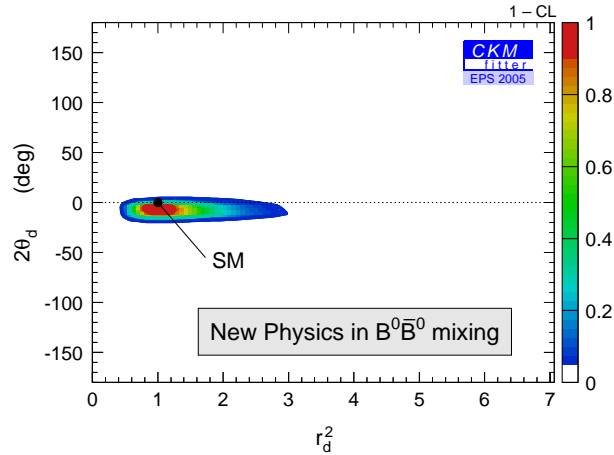}
\caption{Confidence level obtained on the New Physics parameters $r_d$ and $\theta_d$. The preferred region is centred on 
the value corresponding to the Standard Model value ($r_d = 1$ and $\theta_d = 0$).} 
\label{fig:CKMFitter-NP}
\end{center}
\end{figure}

A simple way to test for the presence of New Physics in the \Bd mixing in a model independent way is the following : 
\begin{itemize}
\item Perform a CKM parameters determination using quantities which are involving only tree diagrams, so that
they can be considered as ``New Physics free''. These quantities are $\Vub / \Vcb$ and the information 
on the angle $\gamma$ as obtained from the \BToDK modes \cite{bib:UTFit}.
\item parametrise the presence of New Physics by adding two new parameters $r_d$ and $\theta_d$ : $\Dmd^{\rm{Exp}}= r_d^2 \Dmd^{\rm{SM}}$ 
and ${\cal A}(J/\Psi K^0) = \sin (2 \beta + 2 \theta_d )$,  $\alpha^{\rm{Exp}}= \alpha^{\rm{SM}}-\theta_d$ 
\item Perform the Unitarity Triangle fit~\cite{bib:CKMFitter}, \cite{bib:UTFit} with these extra parameters using all 
available measurements~\footnote{Including the CP asymmetry from the $B$ semi-leptonic decays. 
This measurement of CP violation in the mixing, compatible with 0~\cite{bib:HFAG}, puts strong constraints on $\theta_d$.} 
\end{itemize}
An example of the resulting constraints  on  $r_d$ and $\theta_d$ is shown in Figure~\ref{fig:CKMFitter-NP}. 
The large theoretical uncertainty on the non perturbative QCD parameter $f_{B_d} \sqrt{\hat B_{B_d}}$ which is entering
in the extraction of the CKM matrix element from the \Dmd measurement explains the fact that despite precise measurements
$r_d$ is only poorly constrained. The situation is quite different for $\theta_d$ : the fit selects   $\theta_d \sim 0 $
which indicates that New Physics CP violating phase should be close to the Standard Model one.  

\section{Conclusion}
Charm and beauty physics are entering the precision era. The non perturbative QCD parameter $f_D$ 
is now precisely measured by CLEO-c and is found to be in good agreement with the latest LQCD 
computations. The $B \ra \tau \nu$ decay should be measured within the next years, providing a measurement 
of $f_B$.  
For the first time $b \ra d \gamma$ type decays have been observed, the measurements are not yet precise enough 
but in the forthcoming years , the ratio of $B \ra \rho \gamma$ to $B \ra K^* \gamma$ may provide constraints 
on $\Vtd/\Vts$ complementary to the mixing measurements. 
One of the missing measurements is the \Bs mixing frequency parameter \Dms, if it is not achieved at 
the Tevatron this will be done at the LHC. 
The \Vub measurement using semi-leptonic $B$ decays is now getting quite precise. The inclusive method has
reached a precision of 8\%, the exclusive one is at the limit of being able to distinguish between theoretical models.  
CP violation is measured at 5\% in the charmonium modes, unfortunately one is not yet able to conclude on 
the presence or not of New Physics in the $b \ra s \overline s s $-type decays, more statistics are needed. 
The angle $\alpha$ is now measured with a precision of about 10 \%  using the $\Bz \ra \rho \rho $ decay. 
A precision measurement of the angle $\gamma$ will require more data due to the rather small value of the $r_B$ parameter. 
All these measurements tell us that the Standard Model is an excellent description of CP violating and FCNC processes. 
New Physics in the \Bd mixing seems to have a CP violation phase close to the Standard Model one. The \Bs window on 
new Physics has still to be looked at, this will be one of the most important task of the LHCb experiment at CERN. 
\section{Acknowledgements}
I would like to thank the organisers of this very interesting conference for their invitation in Lisboa. 
I would also like to thank my BABAR colleagues for their support and help in preparing this talk. 
Many thanks to the ``UT-Fitter'' Maurizio Pierini and the ``CKM-Fitter'' Heiko Lacker for providing me 
all the results of their programs. 
I am really indebted to Achille Stocchi for the enlighting discussions we had while preparing this talk 
and for his continuous support.

\end{document}